\documentclass[a4paper,singlecoulomb,12pt]{article}
\usepackage{epsfig,amsmath,amssymb}
\usepackage[colorlinks=true,linktocpage=true,linkcolor=blue,citecolor=blue]{hyperref}
\usepackage{cite}
\usepackage{appendix}
\usepackage{graphicx}
\usepackage{dcolumn}
\usepackage{bm}
\usepackage{subfig}
\usepackage{float}
\usepackage{tabularx}
\usepackage{hhline}
\usepackage{color}
\usepackage{latexsym}
\newcommand{\ba}{\begin{eqnarray}}
\newcommand{\ea}{\end{eqnarray}}

\usepackage{caption}

\begin{document}
\title {\large  {\bf  Transport  coefficients 
  in thermal QCD: A probe to the collision integral }}

\author{Salman Ahamad Khan\footnote{skhan@ph.iitr.ac.in}  and  
Binoy Krishna Patra\footnote{binoy@ph.iitr.ac.in} \vspace{0.3in} \\
Department of Physics, \\
Indian Institute of Technology Roorkee, Roorkee 247667, India}
\date{}
\maketitle
\vskip 0.01in

\begin{flushright}
{\normalsize
}
\end{flushright}

\begin{abstract} 
We have studied the transport coefficients as a tool to probe the 
collision integral appeared in the Boltzmann transport equation. For 
this purpose, we have estimated the transport coefficients 
(momentum: \{$\eta$, $\zeta$\}, heat: \{$\kappa$\}, and charge: 
\{$\sigma_{\rm el}$\}) in the kinetic theory approach with the collision 
integrals in Bhatnagar-Gross-Krook (BGK) and relaxation-time (RT)
approximations and ask whether we can distinguish 
between the two collision integrals. For example, $\eta$ gets enhanced 
while $\zeta$  gets reduced {\em w.r.t.} to RT. As a corollary, we then 
investigate the interplay among the aforesaid transport coefficients, 
{\em viz.} fluidity and transition point of QCD medium by evaluating 
the ratios, $\eta/s$ and $\zeta/s$, respectively, nature of flow 
(Reynolds number, RI), sound attenuation (Prandtl number, Pr),  and 
competition between the momentum and charge diffusion ($\gamma$) etc.
as further plausible tools to decipher the same.  With BGK collision 
integral, the ratios, $\eta/s$ (increase) and $\zeta/s$ (decrease) 
show opposite behavior whereas Pr, RI, $\gamma$  and the ratio,
$\zeta/\eta$ get reduced w.r.t RT.  We then examine how a strong 
magnetic field modulate the impact of collision integral, which, 
in a way, explore the dimensionality dependence of the transport
phenomena, {\em especially} momentum transport because the quark dynamics 
is effectively restricted to 1-D only and 
only the lowest Landau levels are populated.  As a result, $\eta$ 
($\zeta$) gets reduced (amplified), which will have ramifications on 
the ratios, {\em viz.} $\eta/s$ ($\zeta/s$) becomes smaller (larger), 
enhancement of Pr, $\gamma$ and $\zeta/\eta$ etc. In this study, the 
thermo-magnetic medium effects have been incorporated by adopting a 
thermodynamically consistent quasi-particle model, where the medium 
generated masses of the partons have been obtained from the pole of 
their resummed propagators calculated using perturbative thermal QCD 
in strong $B$. 
  \end{abstract}
\section{Introduction}
Heavy ion 
collision experiments at Relativistic Heavy Ion collider (RHIC) 
and Large Hadron Collider (LHC) produce an extremely hot 
and dense phase of matter, dubbed as Quark Gluon Plasma 
(QGP) which has also cosmological and astrophysical relevance. Recent 
investigations at RHIC and LHC conclude that a very intense magnetic 
field (of the order of $m_\pi^2$ at RHIC \cite{Kharzeev:NPA803'2008} 
and $15m_\pi^2$ at LHC \cite{Skokov:IJMPA24'2009}) perpendicular to 
the reaction plane is  also originated due to the non-centrality of 
the collisions. Earlier, it was believed that
the life-time of this strong magnetic field is too short to have any 
observable effect on the bulk properties of the QGP, but later it 
is found that the lifetime gets elongated due to finite 
electrical conductivity. The heavy-ion community pays much attention 
to incorporate the effects of magnetic field on the bulk properties of 
the hot and dense QCD medium, {\em viz.} 
thermodynamical~\cite{Rath:JHEP1712'2017,Karmakar:PRD99'2019}, 
transport 
properties~\cite{Dey:PRD102'2020,Pushpa:PRD105'2022,
Hattori:PRD94'016,Hattori:PRD95'2017,
Harutyunyan:PRC94'2016,Das:PRD101'2020,
Bandyopadhyay:PRD102'2020,
Rath:PRD102'2020,Kurian:EPJC79'2019,
Li:PRD97'2018,
Nam:PRD87'2013,Tuchin:JPG39'2012,Denicol:PRD98'2018,
Ghosh:PRD100'2019,Chen:PRD101'2020,
Huang:PRD81'2010,Huang:AnP326'2011,Hattori:PRD96'2017,
Agasian:PAN76'2013,Agasian:JETP95'2012}, production
of soft photons~\cite{Basar:PRL109'2012},
 dilepton production rate\cite
{Tuchin:PRC88'2013,Bandyopadhyay:PRD94'2016}, refractive 
indices~\cite{Fayazbakhsh:PRD86'2012},
heavy quarkonia~\cite{hasan:NPA995'2020,
hasan:PRD102'2020,Khan:arXiv:2108.12700,singh:PRD97'2018}
and heavy quark dynamics~\cite{Fukushima:PRD93'2016,
Sodofyev:PRD93'2016,Aritra:PRD105'2022}. 
Similarly, the study of salient features in QCD, {\em viz.} 
chiral magnetic 
effect~\cite{Fukushima:PRD78'2008,Kharzeev:NPA803'2008}, 
(inverse) magnetic 
catalysis\cite{Gusynin:PRL73'1994,Gusynin:PRD56'1997},
 axial magnetic 
effect~\cite{Braguta:PRD89'2014,Chernodub:PRB'2014}, 
chiral vortical effect 
\cite{Kharzeev:PRL106'2011, Kharzeev:PPNP88'2016} have been 
resurrected due to the feasibility of strong magnetic field.

Recently we have studied the charge 
and heat transport coefficients 
in an ambience of strong magnetic 
field in kinetic theory framework 
with the collision terms in BGK and relaxation-time (RT) 
approximations~\cite{Khan:PRD104'2021}. We found 
that the BGK and RT collision 
terms yield different solutions to the 
transport equation, which, in 
turn, yield different predictions for
 the heat and charge transport 
coefficients. Since the collision integral 
directly controls the 
momentum distribution of the particles 
so the collision integral could 
have direct impact on the momentum transport coefficients. This 
motivates us to explore the sensitivity of collision terms on the 
momentum transport because the transport coefficients: shear ($\eta$) 
and bulk ($\zeta$) viscosities are very important input parameters 
to model the hydrodynamical evolution of the strongly interacting matter. 
Certain (conformal) field theories, which are dual to gravity in
higher space-time dimensions, show a lower bound 
$1/4\pi$ for the ratio, $\eta/s$ ($s$ is the entropy density)~\cite{
Kovtun:PRL94'2005} and also conjectures this same value as a lower limit
for all substances. RHIC and LHC data, represented by radial and 
elliptic flow, also indicates very small 
viscosity~\cite{Abelev:PRC77'2008,Luzum:PRC78'2008,
Adam:PRL117'2016,Adam:PRL116'2016,
Abelev:PRL111'2013,Gavin:PRL97'2006}.
Lattice simulations also predicts the small
shear viscosity~\cite{Nakamura:PRL94'2005}. 
On the other hand at the classical level, the bulk
viscosity vanishes for the conformal fluid and massless QGP but quantum
corrections break the conformal symmetry and it acquires non-zero value  
even for the massless QGP found in the lattice studies
of $SU(3)$ gauge theory~\cite{Meyer:PRL100'2008}. 
In some other studies, the value of the 
ratio $\eta/s$ is found to be minimum~
\cite{Csernai:PRL97'2006,Lacey:PRL98'2007} 
near the phase-transition 
while the bulk viscosity to be maximum~\cite{Karsch:PLB217'2008,Paech:PRC74'2006}. 
Keeping these studies in the mind, many groups in heavy ion
community have calculated the shear and 
bulk viscosities~\cite{Danielewicz:PRD31'1985,Sasaki:PRC79'2009,
Arnold:JHEP11'2000,Arnold:PRD74'2006,Hidaka:PRD78'2008,
Bernhard:NP15'2019} and some derived coefficients, {\em viz.} Prandtl number~\cite{Schafer:RPP72'2009,
Rangamani:JHEP01'2009,Braby:PRA82'2010},
Reynolds number~\cite{Csernai:PRC85'2012,McInnes:NPB21'2017},
 dimensionless ratio $\gamma=(\eta/s)/(\sigma_{el}/T)$ 
(where $\sigma_{el}$ is the electrical conductivity)
of the hot QCD medium in the kinetic theory~\cite{Thakur:PRD95'2017,
Rath:PRD102'2020}
with RT collision term and Chapman-Enskog approach with quasiparticle description
through a effective fugacity model~\cite{Mitra:PRD96'2017}. 
The relative significance of
  the shear and bulk viscosity ($\zeta/\eta$) is studied in many systems, 
{\em such as}  interacting scalar field~
 \cite{Horsley:NPB280'1987}, hot QCD medium using the perturbation 
theory~\cite{Arnold:PRD74'2006}, strongly coupled gauge 
plasma using the gauge gravity duality~\cite{Buchel:NPB820'2009,
Gursoy:JHEP0912'2009} and for the quasi-gluon plasma~\cite{Bluhm:PLB709'2012}.

\par
The transport coefficients discussed so 
far are limited to thermal QCD
medium in the absence of magnetic field,
 however, the advent of strong 
magnetic field created at RHIC and/or 
LHC motivates researchers to 
estimate the transport coefficients 
in this kind of environment
 because the dissipative magneto-hydrodynamics 
 needs a understanding of the viscosity coefficients in the presence
of magnetic field.
In strong $\vec{B}$, only the lowest Landau levels (LLLs) are 
populated, which in turn constrains the dynamics of quarks to 1-D. 
This dimensional reduction in phase space has deep impact 
on the collision integral (through the relaxation time), which subsequently 
modifies the solution 
of Boltzmann equation. However, the strong $B$ modulates the transport 
coefficients directly because the tensorial structure of the transport
coefficients gets modified.
In case of momentum transport, the viscous tensor in $\vec{B}$ 
have seven independent tensors, compared to only two 
independent tensors in absence of
magnetic field. That is, the number
of viscosity coefficients in a magnetic field are 
seven, out of which five are 
shear, one is bulk and  the last one is the cross term between
ordinary and bulk viscosities. When the strength of $\vec{B}$ is very strong,
the transverse components of velocity
(w.r.t. the direction of $\vec{B}$)
vanish, so non-diagonal terms vanish and only diagonal (longitudinal) 
terms survive. The nonvanishing terms are further 
grouped into traceless and nonzero trace tensor, the coefficients
of them are (longitudinal) shear and 
bulk viscosities, respectively.
These coefficients in magnetic field 
have been studied recently in kinetic 
theory~\cite{Rath:PRD102'2020} with RT collision term, Chapman Enskog method~
\cite{Kurian:EPJC79'2019}, perturbative QCD in weak magnetic field~
\cite{Li:PRD97'2018}, Kubo formalism~\cite{Hattori:PRD96'2017,Chen:PRD101'2020,
Li:PRD97'2018,Nam:PRD87'2013} and 
using the holographic technique
\cite{Jain:JHEP10;2015,Finazzo:PRD94'2016}. 
Apart from the influences of $\vec{B}$, and
collision integral,
another important factors in transport phenomena is the quasiparticle 
descriptions for partons.

The aim of the present paper is to treat 
the transport coefficients as
a probe to understand the microscopic
 aspects of the medium in three
fold respects: (i) 
By looking into the relative behaviour 
of transport coefficients
(shear and bulk viscosities, thermal 
and electrical conductivities etc.) 
due to the collision integral of different kinds used in the transport
equation, we can decipher the nature of collision term.
Here we use BGK collision term and the commonly used simplified RT collision 
term. (ii) Once we decipher the correspondence between
the transport coefficients and the collision integral, we will further
ask whether this correspondence is still 
restored in dimensional reduction
(3 $\rightarrow$ 1 dimension), which is made 
 possible in a strong magnetic
field.  (iii) We explore the aforesaid correspondence from higher to
lower dimension (artifact of strong $\vec{B}$) in an realistic description 
of partons in thermal QCD medium, known as quasiparticle model (QPD). 
A hot and dense QCD medium manifests a hierarchy in mass scales
perturbatively: $T \ll gT \ll g^2 T$ ($g$ is strong running coupling constant).
Physically, $gT$ is the (electric screening) scale 
at which the system 
develops collective oscillation, which is also 
viewed as the mass generated
to the partons in a thermal medium. We envisage  QPD
by assigning masses to partons, which are obtained from 
their dispersion relations (datails are in 
Section 2)~\cite{Bannur:JHEP09'2007}. 
There are many versions of 
QPDs, {\em such as} effective theories based models: 
Nambu-Jona-Lasinio (NJL) and PNJL models
\cite{Fukushima:PLB591'2004,
Ghosh:PRD73'2006,Abuki:PLB676'2009} models, 
effective fugacity model~\cite{Chandra:PRD84'2011},
models based on the Gribov–Zwanziger quantization
~\cite{Su:PRL114'2015,Florkowski:PRC94'2016,
Jaiswal:PLB11'2020} etc.

The BGK collision integral in the transport equation has been used 
earlier in the calculation of the refractive index~\cite{Jiang:PRD94'2016}, 
the dielectric permitivity~\cite{Carrington:CJP82'2004}, heavy quark energy 
 loss~\cite{Yousuf:EPJC79'2019} and  collective modes of the 
  hot QCD medium~\cite{Kumar:PRD97'2018}.
In this work, we have calculated the shear and bulk viscosities
in kinetic theory with quasiparticle description of partons. In addition 
to viscosity coefficients, we have revisited 
our recent calculations~\cite{Khan:PRD104'2021}
on thermal and electrical conductivities, to study the impact of 
collision terms on the specific viscosities ($\eta/s$, $\zeta/s$), Prandtl 
number, Reynolds number, dimensionless ratio $\gamma$, and the ratio
 $\zeta/\eta$ etc. The entire aforesaid calculations are done 
in two steps: in the absence and presence of strong magnetic field. In
$\vec{B}=0$ case, both $\eta$ and $\zeta$ increases with the temperature 
. The magnitude of $\eta$ ($\zeta$) gets enhanced (reduced)
 in BGK collision term in comparison to RT, wherein $\eta/s$ is minimum
while $\zeta/s$ is maximum near the deconfinement temperature ($T_C \sim 0.16
$ GeV). The Prandtl and Reynolds numbers 
have been found to be increasing with $T$.
The magnitude of the  Pr, RI, $\gamma$ and $\zeta/\eta$ slightly get 
reduced in BGK collision term. 
In the presence of strong $\vec{B}$, the abovementioned transport
coefficients show similar trends {\em w.r.t } the collision
integral except $\zeta/s$ which has almost same 
magnitude in both the collision integrals. 
The coefficients, Pr, $\gamma$ and
$\zeta/\eta$ become larger while RI becomes smaller. 
Thus, we observe that 
the collision integrals of different kinds predict 
different values of the 
transport coefficients. Transport coefficients are 
experimentally measurable quantities whereas collision
integral is a theoretical input to the transport (Boltzmann) equation,
thus by inverting this one-to-one correspondence, we 
can have an idea about the collision integral
with the help of the transport coefficients. 
 
The manuscript has been organized in the following manner:
We have started with the discussion on a hot QCD medium in an ambience
of strong magnetic field in Section~\ref{2}, whose dispersion relation
gives rise the quasi-particle dispersion of partons through the 
masses generated by the artifact of a thermal medium.
In section~\ref{3}, we have discussed the general framework of the kinetic 
theory approach
and collision integrals to linearize the Boltzmann transport equation.
In subsections~\ref{3.1} and~\ref{3.2}, we calculate 
the shear and bulk viscosities in absence 
as well as in the strong $B$, respectively. 
In subsection~\ref{3.3}, we 
have
revisited our earlier study related to the 
charge and heat transport. Further in section~\ref{4}, we discuss the 
various derived coefficients {\em like} $\eta/s$, $\zeta/s$
 Pr, RI, $\gamma$ factor and at last the ratio $\zeta/\eta$.
Finally, we conclude 
in section~\ref{5}.

\section{Quasiparticle Model}\label{2}
 The basic
idea behind  quasi-particle models is that one
can study the various properties of a system
of strongly interacting quarks and gluon by considering it
as an ensemble of  quasi-quarks 
and quasi-gluons, where the information about the
interaction is hidden in the
medium generated mass of the quasi-particles.
 The quasi-particles manifest a collective 
behavior of the medium, not the independent 
singular nature of a parton.  
 There are many  effective models in the 
literature  to study  the various properties
of the QGP~\cite{Bannur:JHEP09'2007,Fukushima:PLB591'2004,
Ghosh:PRD73'2006,Abuki:PLB676'2009,Chandra:PRD84'2011,
Su:PRL114'2015,
 Florkowski:PRC94'2016,Jaiswal:PLB11'2020}.
In the present work, we have exploited a thermodynamically 
consistent  phenomenological
model~\cite{Bannur:JHEP09'2007} which explains the 
lattice QCD data well
in its domain of applicability. In this model, 
the quasi-particle mass
of the quark has been parameterized as
\begin{eqnarray}
m_i^2=m_{i0}^2+\sqrt{2}m_{i0}m_{iT}+m_{iT}^2,
\label{para_massT}
\end{eqnarray}
where $m_{i,0}$ and $m_{i,T}$ are the current quark  and the 
medium generated masses of the $i^{th}$ quark flavor, respectively. The 
medium generated thermal masses of the quarks have been 
calculated by using the perturbative thermal QCD in the
 HTL approximation~\cite{Braaten:PRD45'1992,Peshier:PRD66'2002}
\begin{eqnarray}
m_{iT}^2=\frac{g'^2T^2}{6},
\label{quarkmassT}
\end{eqnarray}
where $g' (=\sqrt{4\pi\alpha'_s})$ is the QCD coupling constant
 which depends on the 
temperature and $\alpha'_s$  
is given by its one-loop expression:
 \begin{eqnarray}
\alpha'_s (T)=\frac{6\pi}{(33-2N_f) \ln\left(\frac{Q}{\Lambda_{QCD}}\right)},
\label{coupling_T}
\end{eqnarray} 
the scale, $Q$ is set at $2\pi T$ ($T$ is the temperature).\\

Similarly, gluons 
also acquire a thermal mass
as a result of the interaction with the surrounding partons
in the HTL approximation as~\cite{Peshier:PRD66'2002,Bellac:1996}
\ba 
m_g^2=\frac{g'^2T^2}{6}\left(N_C+\frac{N_f}{2}\right).
\label{gluonmassT} 
\ea
In the presence of a strong magnetic field,
we will use a similar parameterized mass for the quasi-particles 
\begin{eqnarray}
m_i^2=m_{i0}^2+\sqrt{2}m_{i0}m_{iT, B}+m_{iT, B}^2,
\label{para_massTB}
\end{eqnarray}
where  $m_{iT,B}$ is the thermally generated mass which is calculated 
from the poles of the resummed quark  propagator in the strong 
magnetic field using the self-consistent  Dyson-Schwinger equation 
\begin{eqnarray}
S^{-1}(p_{\parallel})=\gamma^{\mu}p_{\parallel \mu}-\Sigma(p_{\parallel}).
\label{dyson}
\end{eqnarray}
 $\Sigma(p_{\parallel})$ in the above Eqn.~\eqref{dyson} refers to the 
quark self-energy which can be written up to one loop in 
presence of temperature and  strong 
$B$ as 
\begin{eqnarray}
\Sigma(p)=-\frac{4}{3} g^{2}i\int{\frac{d^4k}{(2\pi)^4}}
\left[\gamma_\mu {S(k)}\gamma_\nu {D^{\mu \nu} (p-k)}\right],
\label{quark_self}
\end{eqnarray}
where $4/3$ corresponds to the Casimir factor and
the strong coupling  $g$ now runs with the magnetic field only~\cite{Ferrer:PRD91'2015}
\begin{eqnarray}
\frac{g^2}{4\pi}=\alpha_s(|q_fB|)=\frac{1}{(\alpha^0(\mu_0))
^{-1}+\frac{11N_C}{12\pi}
\ln \left(\frac{{\rm k_z}^2+M_B^2}{\mu_0^2}\right)+\frac{1}{3\pi}
\sum_i \frac{|q_iB|}{\sigma}},
\label{coupling_B}
\end{eqnarray}
where  $$\alpha^0(\mu_0)=\frac{12\pi}{11N_C \ln
\left(\frac{\mu_0^2+M_B^2}{\Lambda_V^2}\right)}.$$
Here $M_B$ refers to the infrared mass (1 GeV) and $\Lambda_V$ and $\mu_0$
are chosen as 0.385 GeV and 1.1 GeV, respectively and ${\rm k_z}
=0.1\sqrt{eB}$.\par
$S(k)$ corresponds to the quark propagator in an external magnetic
field which has been calculated by the
Schwinger proper-time method~\cite{Schwinger:PR82'1951} as
\begin{equation}
iS (k)=\sum_n\frac{-id_n(\alpha)D+d^\prime_n(\alpha)
\bar{D}}{k_\|^2 -m^2_{i,0}+2n|q_fB|}
+i\frac{\gamma\cdot k_\bot}{k_\bot^2}~,
\label{prop_lag}
\end{equation}
where the variable $\alpha$ = $k_\bot^2/|q_fB|$ is dimensionless and 
$D$, $\bar{D}$, $d_n(\alpha)$ and $d'_n(\alpha)$ read as~\cite{Chyi:PRD62'2000} 
\begin{eqnarray*}
&&D=(m_{i,0}+\gamma\cdot k_\|)+\gamma\cdot
k_\bot\frac{m_{i,0}^2-k_\|^2}{k_\|^2},\\
&& \bar{D}=\gamma_{1}\gamma_{2}(m_{i,0}+\gamma\cdot k_\|),\\
&& d_n(\alpha)=(-1)^n e^{-\alpha}C_n(2\alpha),\\
&& d^{'}_{n}(\alpha)=\frac{\partial d_n}{\partial\alpha}.
\end{eqnarray*}
The $C_n$'s can further be expressed in terms of Laguerre polynomial ($L_n$)
\begin{eqnarray*}
C_n(2\alpha)&=&L_{n}(2\alpha)-L_{n-1}(2\alpha).
\end{eqnarray*}
The above quark propagator $S(k)$ is simplified in  the 
 strong magnetic field ($|q_fB| >> T^2$) regime as 
 \ba\label{quark_prop}
 S(k)=ie^{-\frac{k^2_\perp}{|q_iB|}}
 \frac{\left(\gamma^0 k_0-\gamma^3 k_z+m_{i,0}
 \right)}{k^2_\parallel-m^2_{i,0}}\left(1
 -\gamma^0\gamma^3\gamma^5\right)
 ,\ea
 where the perpendicular and parallel part of the 
  metric tensors are defined as
\begin{eqnarray*}
g^{\mu\nu}_\perp &=&{\rm{diag}}(0,-1,-1,0), \\
g^{\mu\nu}_\parallel &=& {\rm{diag}}(1,0,0,-1),
\end{eqnarray*}
similarly of four vectors as
\begin{eqnarray*}
 k_{\perp\mu} \equiv(0,k_x,k_y,0),\\
   k_{\parallel\mu}\equiv(k_0,0,0,k_z).
\end{eqnarray*} 
$D^{\mu \nu}(p-k)$ corresponds to the gluon propagator which is 
 not affected by strong $B$
 \ba
 \label{g. propagator}
 D^{\mu \nu} (p-k)=\frac{ig^{\mu \nu}}{(p-k)^2}.
\ea
The quark self-energy~\eqref{quark_self} can be further 
simplified using the imaginary-time formalism as
(see the Appendix B)\cite{Rath:PRD100'2019}
 \begin{eqnarray}
\Sigma(p_\parallel)=\frac{g^2|q_iB|}
{3\pi^2}\left[\frac{\pi T}{2m_{i,0}}-\ln(2)\right]
\left[\frac{\gamma^0p_0}{p_\parallel^2}+\frac{\gamma^3p_z}{p_\parallel^2}+\frac{\gamma^0\gamma^5p_z}{p_\parallel^2}+
\frac{\gamma^3\gamma^5p_0}{p_\parallel^2}\right].
\end{eqnarray}
The general structure of the quark self energy in
the strong magnetic field can be written as
\cite{Ayala:PRD91'2015,Karmakar:PRD99'2019}
\begin{eqnarray}
\Sigma(p_{\parallel})=A\gamma^{\mu}u_{\mu}+B\gamma^{\mu}b_{\mu}
+C\gamma^{5}\gamma^{\mu}u_{\mu}+D\gamma^{5}\gamma^{\mu}b_{\mu},
\label{form_factor}
\end{eqnarray}
 where $u^{\mu}(1,0,0,0)$ 
and $b^{\mu}(0,0,0,-1)$ correspond to  
the direction of the heat bath and magnetic field,
 respectively. The form factors A, B,
C, and D  are calculated 
 in LLL as
\begin{eqnarray}
A&=&\frac{1}{4}{\rm Tr}[\Sigma\gamma^{\mu}u_{\mu}]=
\frac{g^2|q_iB|}{3\pi^2}\left[\frac{\pi T}{2m_{i,0}}-\ln{(2)}\right]
\frac{p_0}{p_{\parallel}^2},\\
B&=&-\frac{1}{4}{\rm Tr}[\Sigma\gamma^{\mu}b_{\mu}]=
\frac{g^2|q_iB|}{3\pi^2}\left[\frac{\pi T}{2m_{i,0}}-\ln{(2)}\right]
\frac{p_z}{p_{\parallel}^2},\\
C&=&\frac{1}{4}{\rm Tr}[\gamma^5 \Sigma\gamma^{\mu}u_{\mu}]=-
\frac{g^2|q_iB|}{3\pi^2}\left[\frac{\pi T}{2m_{i,0}}-\ln{(2)}\right]
\frac{p_z}{p_{\parallel}^2},\\
D&=&-\frac{1}{4}{\rm Tr}[\gamma^5\Sigma\gamma^{\mu}b_{\mu}]=-
\frac{g^2|q_iB|}{3\pi^2}\left[\frac{\pi T}{2m_{i,0}}-\ln{(2)}\right]
\frac{p_0}{p_{\parallel}^2}.
\end{eqnarray}
 We observe from the above equations that $C=-B$ and $D=-A$. \par
 Using the right and left hand 
 chiral projector operators 
 $P_R=(1+\gamma^{5})/2$ and $P_L=(1-\gamma^{5})/2$, 
 we can express the self energy~\eqref{form_factor} as
\begin{eqnarray}
\Sigma(p_{\parallel})=P_R[(A-B)\gamma^{\mu}u_{\mu}+(B-A)\gamma^{\mu}b_{\mu}]P_L
+P_L[(A+B)\gamma^{\mu}u_{\mu}+(B+A)\gamma^{\mu}b_{\mu}]P_R.
\end{eqnarray}
 The inverse effective quark propagator~\eqref{dyson}
 can be written in terms of the chiral projector operators as 
\begin{eqnarray}
S^{-1}(p_{\parallel})=P_R\gamma^{\mu}X_{\mu}P_L+P_L\gamma^{\mu}Y_{\mu}P_R,
\end{eqnarray}
where 
\begin{eqnarray}
\gamma^{\mu}X_{\mu}&=&\gamma^{\mu}p_{\parallel \mu}-
(A-B)\gamma^{\mu}u_{\mu}-(B-A)\gamma^{\mu}b_{\mu},\\
\gamma^{\mu}Y_{\mu}&=&\gamma^{\mu}p_{\parallel \mu}-
(A+B)\gamma^{\mu}u_{\mu}-(B+A)\gamma^{\mu}b_{\mu}.
\end{eqnarray}
The effective propagator can be further written as
\begin{eqnarray}
S(p_{\parallel})=\frac{1}{2}\left[P_R\frac{\gamma^{\mu}Y_{\mu}}{Y^2/2}P_L
+P_L\frac{\gamma^{\mu}X_{\mu}}{X^2/2}P_R\right],
\end{eqnarray}
where
\begin{eqnarray}
&&\frac{X^2}{2}=X_1^2=\frac{1}{2}\left[p_0-(A-B)\right]^2-
\frac{1}{2}\left[p_z+(B-A)\right]^2 ~, \\
&&\frac{Y^2}{2}=Y_1^2=\frac{1}{2}\left[p_0-(A+B)\right]^2-
\frac{1}{2}\left[p_z+(B+A)\right]^2
~.\end{eqnarray}
In order to get the medium generated mass, 
we take static limit ($p_0 = 0,~p_z \rightarrow 0$)  
of either $X_1^2$ or $Y_1^2$ (which are equal in LLL). 
The thermal mass has been obtained as
\begin{eqnarray}
m_{iT,B}^2=\frac{g^2|q_iB|}{3\pi^2}\left[\frac{\pi T}
{2m_{i,0}}-\ln{(2)}\right],
\label{massTB}
\end{eqnarray}
which exhibits dependence on both $B$ and $T$. 
The strong magnetic field does not have any direct impact on the 
gluons rather, the quark loop's contribution 
to  it's self-energy gets modified.  
 The thermal mass of the gluons in the 
presence of the strong $B$ has been 
calculated as\cite{fukushima:PRD93'2016,singh:PRD97'2018}
\ba 
m_{g,B}^2=\frac{g'^2T^2N_C}{6}
+\sum_f \frac{g^2|q_fB|}{8\pi^2}.
\label{gluonmassB} 
\ea
We have exploited these medium generated masses in  
distribution functions  of the quarks and gluons to evaluate the transport
coefficients in the next section.
\section{Momentum transport in a thermal medium of  
quarks and gluons} \label{3}
In the kinetic theory approach, the time 
evolution of the plasma system is governed by the 
relativistic Boltzmann transport equation (RBTE),
 which is given for a
single particle distribution function as
\begin{eqnarray}
p^{\mu}\partial_{\mu} f(x,p)= C[f(x,p)],
\label{bte}
\end{eqnarray}
 where $C[f(x,p)]$ is the collision integral which encodes 
the collision processes in the medium.  
In a realistic scenario, we need  transition rates of the various QCD 
processes to get the collision integral $C[f]$, but for a qualitative 
analysis of the transport phenomenon, $ C[f]$ can be written by using 
the mean free path treatment in RT approximation 
 \ba
 C[f]=-\frac{p^\mu u_\mu}{\tau} 
 \left(f(\vec{x},\vec{p},t)-f_{\rm eq} (|\vec{p}|)
\right),
\label{RT}
\ea
where $\tau (p)$ is the time period 
after which the distribution 
function attains the equilibrium 
configuration $f_{\rm eq} (|\vec{p}|)$.  
 This time  $\tau (p)$  is a free 
 parameter commonly  known as 
 {\em relaxation time}. The problem 
 with RT collision term is that it violates the 
 charge and particle number
 conservation.  
Later, the RT collision term was 
  modified to ensure the particle number and charge
conservation by   
  Bhatnagar-Gross-Krook (BGK) as ~\cite{Bhatnagar:PRD94'1954,Schenke:PRD73'2006}   
 \begin{eqnarray}
C[f] = -\frac{p^\mu u_\mu}{\tau} 
\left(f(\vec{x},\vec{p},t)- \frac{n(\vec{x},t)}
{n_{\mathrm eq}} f_{\rm eq} (|\vec{p}|)\right),
\label{bgk_col}
\end{eqnarray}
which conserves the particle number instantaneously {\em i.e.}
\begin{eqnarray}
\int \frac{d^3p}{(2\pi)^3} C[f]=0.
\end{eqnarray}
 Therefore, BGK collision term is found to
yield the simple minded RT term in a special case, when the 
instantaneous
number density, $n(\vec{x},t)$ of the system 
during the deviation from its
equilibrium becomes equal to the equilibrium
density, $n_{\rm eq} (\vec{x},t)$. Since $n(\vec{x},t)/n_{\rm eq} 
(\vec{x},t) < 1 $ so the magnitude of BGK collision term is more than
RT term, which could have varied ramifications on the transport coefficients.\\

Now in the forthcoming sections, we  will investigate the effects of the 
BGK collision term on the  momentum transport in the thermal QCD medium 
and will further see the implications of the strong magnetic field on 
BGK collision term, which will indirectly modulate the 
momentum transport through the relaxation time,  
phase-space and medium-generated masses of the partons.
\subsection{Shear $(\eta)$ and Bulk $(\zeta)$ viscosities
 in the absence of magnetic field}\label{3.1}
{ Now, we will calculate viscous coefficients $\eta$ and 
$\zeta$ of a thermal QCD medium for 
 $B=0$ case.  We assume that the system is in 
local equilibrium with local temperature $T(x)$ and flow
velocity $u^{\mu}(x)$. Here, $u^{\mu}(x)$ represents the velocity
of baryon number flow in the Eckart frame while the velocity of energy
transport in the Landau-Lifshitz frame. To 
know the response of the system while inflicted with a velocity 
gradient, we allow the system to  
get shifted slightly from equilibrium, the infinitesimal deviation
in the energy-momentum tensor can be written as
\ba
\pi^{\mu \nu}=  T^{\mu \nu}-T_{\rm eq}^{\mu \nu},
\ea
where $T^{\mu \nu}$  refers to 
the energy momentum tensor of the 
system in the out of the equilibrium state, which reads 
\ba
T^{\mu\nu}&=& \int \frac{d^3p}{(2\pi)^3}
{p^{\mu}p^{\nu}}\left[\sum_i 2g_i\frac{f_{i}}
{\omega_i}+g_g\frac{f_{g}}{\omega_g}\right].
\label{EM_noneq}
\ea
Here, factor $2$ in the above Eqn.~\eqref{EM_noneq}
 corresponds to the equal contributions from
quarks and anti-quarks since we are interested in medium with 
zero chemical potential ($\mu =0$). Apart from this,
 $g_i$ and $g_g$ are the quark 
and gluon degeneracy factors, 
respectively. The dissipative part, which 
contains all the necessary information about how  the system   
approaches equilibrium from the non-equilibrium state reads 
\ba
\pi^{\mu\nu}&=& \int \frac{d^3p}{(2\pi)^3}
{p^{\mu}p^{\nu}}\left[\sum_i 2g_i\frac{\delta f_{i}}
{\omega_i}+g_g\frac{\delta f_{g}}{\omega_g}\right],
\label{EM_delta}
\ea
where, $\delta f_i$ ($\delta f_g$) is the infinitesimal deviation
in the equilibrium distribution function for quarks (gluons).
This deviation for the $i^{th}$ flavor reads as
 $\delta f_i=f_i-f_{\mathrm eq, i}$ ,
 where $f_{\rm eq,i}$
is the equilibrium distribution function 
\ba
f_{\mathrm eq, i} = \frac{1}{e^{\beta u^\alpha {p_{\rm i}}_\alpha}+1},
\label{dis_quark}
\ea
where $p_{\rm i}^\alpha$ is $(\omega_i, \mathbf{p})$ 
and $u_\alpha$ refers to the four-velocity of the fluid, 
which can be written as
$u_{\alpha}=(1,0,0,0)$ in case of local rest frame, whereas 
the energy $\omega_i=\sqrt{\mathbf{p}^2+m_i^2}$. Similarly, 
deviation $\delta f_g$ for the gluon is defined as 
$\delta f_{\rm g}=f_{\rm g}-f_{\rm eq,g}$, where the equilibrium 
distribution for gluons reads
\ba
f_{\mathrm eq, g} = \frac{1}{e^{\beta u^\alpha p_\alpha}-1},
\label{dis_gluon}
\ea
and $p_{\alpha}$ is $(\omega_g, \mathbf{p})$.\\

The deviations $\delta f_i$ and $\delta f_g$ will be obtained
from the relativistic linearized Boltzmann transport equation 
with the BGK collision term~\eqref{bgk_col} for quarks and gluons, respectively
\begin{eqnarray} \label{RBTE_quark}
p^{\mu}\partial_{\mu}f_i(x,p)&=&-p^{\mu}u_{\mu}\nu_i 
\left(f_i-n_{i}~n_{\mathrm eq,i}^{-1} 
~f_{\mathrm eq,i}\right),\\
 p^{\mu}\partial_{\mu}f_g(x,p)&=&-p^{\mu}u_{\mu}\nu_g
 \left(f_g-n_{g}n_{\mathrm eq,g}^{-1} f_{\mathrm eq,g}
\right),
\label{RBTE_gluon}
\end{eqnarray}
where
\ba
n_i &=& g_i \int \frac{d^3p}{(2\pi)^3}~ 
(f_{\mathrm eq,i}+\delta f_i),\\
n_{\mathrm eq,i} &=& g_i \int \frac{d^3p}{(2\pi)^3}~
f_{\mathrm eq,i} \label{num_quark}, \\
n_g &=& g_g \int \frac{d^3p}{(2\pi)^3} ~
(f_{\mathrm eq,g}+\delta f_g),\\
n_{\mathrm eq,g} &=& g_g \int \frac{d^3p}{(2\pi)^3}~
f_{\mathrm eq,g}.
\label{num_gluon}
\ea

The collision frequencies of the quarks and 
gluons, $\nu_i$ and $\nu_g$ are 
given by  the inverse of their  respective 
relaxation times, $\tau_i$ 
and $\tau_g$~\cite{Hosoya:NPB250'1985}
\ba
\tau_i(T) &=& \frac{1}{5.1T \alpha'{_s^2} 
\log\left(\frac{1}{\alpha'_s}\right)
[1+0.12(2N_i+1)]}, \nonumber\\  
\tau_g(T) &=& \frac{1}{22.5T \alpha'{_s^2} 
\log\left(\frac{1}{\alpha'_s}\right)
\label{tau_B0} 
[1+0.06N_i]}.  
\ea 
where $\alpha'_s$ is the running coupling 
given by Eqn.~\eqref{coupling_T}.

In nearly equilibrium approximation ($\delta f_i \ll f_{\rm eq, i}$), the 
linearized RBTE  \eqref{RBTE_quark} for ${\rm i^{th}}$ flavour is cast 
into the form\footnote{We use  the notation for momentum
integration, $\int_{p} = \int \frac{d^3p}{(2 \pi)^3}$.} 
(see Appendix A)
\ba 
\delta f_i- g_i~n_{\mathrm eq,i}^{-1}~ 
f_{\mathrm eq,i} \int_{p}\delta f_i&=&
-\frac{\tau_i ~}{p^{\nu}u_{\nu}} p^{\mu}~\partial_{\mu}f_i(x,p),
\label{BGk_left}
\ea
which is then solved iteratively upto first order as
\ba
\delta f_{i}&=&\delta f_i^{(0)}
+ g_i~n_{\mathrm eq,i}^{-1} ~f_{\mathrm eq,i} 
\int_{p'}\delta f_i^{(0)}.
\label{quark_delta}
\ea

The zeroth-order in deviation, $\delta f_i^{(0)}$ is expressed
in terms of temperature and velocity gradients for a medium in
local equilibrium, $T(\vec{x},t)$ with a flow-velocity profile, $\vec{u}
(\vec{x},t)$ for partons:
\ba 
\delta f_i^{(0)}&=&-\frac{\tau_i}{\omega_i}
p^{\mu}~\partial_{\mu}f_i(x,p)\nonumber\\
&=&-\frac{\tau_i ~ }{\omega_iT}p^{\mu}
f_{\mathrm eq,i}(1-f_{\mathrm eq,i})
\left[u_{\alpha}p^{\alpha}u_{\mu}\frac{DT}{T}+
u_{\alpha}p^{\alpha}\frac{\nabla_{\mu}T}{T}-
u_{\mu}p^{\alpha}Du_{\alpha}
-p^{\alpha}\nabla_{\mu}u_{\alpha}\right],
\label{quark_0}
\ea

where the gradients in flow velocity and the temperature can be expressed 
as the sum of the time- and  space-part in the covariant form through 
$D=u^{\mu}\partial_{\mu}$, where $\partial_{\mu}=u_{\mu}D+\nabla_{\mu}$.

Similarly, we evaluate the response by the gluonic component in terms of 
the deviation, $\delta f_g$ from RBTE~\eqref{RBTE_gluon} for 
gluons
\ba
\delta f_{g}&=&\delta f_g^{(0)}
+ g_g~n_{\mathrm eq,g}^{-1} ~f_{\mathrm eq,g} \int_{p'}\delta f_g^{(0)},
\label{gluon_delta}
\ea
where
\ba 
\delta f_g^{(0)}&=&-\frac{\tau_g ~}{\omega_g} 
p^{\mu}~\partial_{\mu}f_g(x,p),\nonumber\\
&=&-\frac{\tau_g ~}{\omega_gT} p^{\mu}
f_{\mathrm eq,g}(1+f_{\mathrm eq,g})
\left[u_{\alpha}p^{\alpha}u_{\mu}\frac{DT}{T}+
u_{\alpha}p^{\alpha}\frac{\nabla_{\mu}T}{T}
-u_{\mu}p^{\alpha}Du_{\alpha}
-p^{\alpha}\nabla_{\mu}u_{\alpha}\right].
\ea

Using the gradients in terms of equation of state
\begin{eqnarray}
\frac{DT}{T} &=& -\left(\frac{\partial P}
 {\partial \varepsilon}\right)\nabla_{\alpha}u^{\alpha}, \nonumber\\
 D u_{\alpha} &=& \frac{\nabla_{\alpha} P}{\varepsilon+P},
\end{eqnarray}
the first-order viscous tensor~\eqref{EM_delta} is thus obtained as
\ba
\pi^{\mu\nu}&=&2\sum_i g_i\int \frac{d^3p}{(2\pi)^3}
\frac{p^{\mu}p^{\nu}}{\omega_i T}\tau_i\bigg\{f_{\mathrm eq,i}
(1-f_{\mathrm eq,i})
\left[\omega_i\left(\frac{\partial P}{\partial \varepsilon}\right)
\nabla_{\alpha}u^{\alpha}
+p^{\alpha} \left(\frac{\nabla_{\alpha}P}{\varepsilon +P}
-\frac{\nabla_{\alpha}T}{T}\right)\right. \nonumber \\ 
&&\left. +\frac{p^{\alpha}p^{\beta}}{\omega_i}\nabla_{\alpha}
u_{\beta}\right]
+g_in_{\mathrm eq,i}^{-1} f_{\mathrm eq,i} 
\int_{p^{\prime}}f_{\mathrm eq,i}
(1-f_{\mathrm eq,i})
\left[\omega '_i\left(\frac{\partial P}{\partial \varepsilon}\right)
\nabla_{\alpha}u^{\alpha}
+p'^{\alpha} \left(\frac{\nabla_{\alpha}P}{\varepsilon +P}
-\frac{\nabla_{\alpha}T}{T}\right)\right. \nonumber\\
&&\left. +\frac{p'^{\alpha}p'^{\beta}}{\omega'_i}\nabla_{\alpha}
u_{\beta}\right]\bigg\}
+ g_g\int \frac{d^3p}{(2\pi)^3}
\frac{p^{\mu}p^{\nu}}{\omega_g T}\tau_g
\bigg\{f_{\mathrm eq,g}(1+f_{\mathrm eq,g})
\left[\omega_g \left(\frac{\partial P}{\partial \varepsilon}\right)
\nabla_{\alpha}u^{\alpha}
+p^{\alpha} \left(\frac{\nabla_{\alpha}P}
{\varepsilon +P}\right. \right. \nonumber\\
&&\left. \left. -\frac{\nabla_{\alpha}T}{T}\right)
 +\frac{p^{\alpha}p^{\beta}}{\omega_g}
\nabla_{\alpha}u_{\beta}\right]
+g_gn_{\mathrm eq,g}^{-1} f_{\mathrm eq,g} 
\int_{p'}f_{\mathrm eq,g}(1+f_{\mathrm eq,g})
\left[\omega'_g \left(\frac{\partial P}
{\partial \varepsilon}\right)
\nabla_{\alpha}u^{\alpha}
+p'^{\alpha} \left(\frac{\nabla_{\alpha}P}
{\varepsilon +P} \right. \right. \nonumber\\
&& \left. \left. -\frac{\nabla_{\alpha}T}{T}\right)
 +\frac{p'^{\alpha}p'^{\beta}}{\omega'_g}\nabla_{\alpha}u_{\beta}\right]\bigg\},
\label{EM_delta1}
\ea
where $\omega'=\omega(p')=\sqrt{p'^2+m^2}$.

Finally, the nonvanishing 
spatial components of $\pi^{\mu \nu}$ tensor yield into the form 
(with the help of definitions: $D\equiv (\partial_t,0)$, 
$\nabla_{\mu}\equiv(0, \partial_i)$) 
\ba
\pi^{ij}&=&2\sum_i g_i\int \frac{d^3p}{(2\pi)^3}
\frac{p^{i}p^{j}}{\omega_i T}\tau_i\bigg\{f_{\mathrm eq,i}
(1-f_{\mathrm eq,i})
\left[\omega_i\left(\frac{\partial P}{\partial \varepsilon}\right)
\nabla_{k}u^{k}
+p^{k} \left(\frac{\nabla_{k}P}{\varepsilon +P}
-\frac{\nabla_{k}T}{T}\right)\right. \nonumber \\ 
&&\left. +\frac{p^{k}p^{l}}{\omega_i}\nabla_{k}
u_{l}\right]
+g_in_{\mathrm eq,i}^{-1} f_{\mathrm eq,i} 
\int_{p^{\prime}}f_{\mathrm eq,i}
(1-f_{\mathrm eq,i})
\left[\omega '_i\left(\frac{\partial P}{\partial \varepsilon}\right)
\nabla_{k}u^{k}
+p'^{k} \left(\frac{\nabla_{k}P}{\varepsilon +P}
-\frac{\nabla_{k}T}{T}\right)\right. \nonumber\\
&&\left. +\frac{p'^{k}p'^{l}}{\omega'_i}\nabla_{k}
u_{l}\right]\bigg\}
+ g_g\int \frac{d^3p}{(2\pi)^3}
\frac{p^{i}p^{j}}{\omega_g T}\tau_g
\bigg\{f_{\mathrm eq,g}(1+f_{\mathrm eq,g})
\left[\omega_g \left(\frac{\partial P}{\partial \varepsilon}\right)
\nabla_{k}u^{k}
+p^{k} \left(\frac{\nabla_{k}P}
{\varepsilon +P} \right. \right. \nonumber\\
&& \left. \left. -\frac{\nabla_{k}T}{T}\right)
 +\frac{p^{k}p^{l}}{\omega_g}
\nabla_{k}u_{l}\right]
+g_gn_{\mathrm eq,g}^{-1} f_{\mathrm eq,g} 
\int_{p'}f_{\mathrm eq,g}(1+f_{\mathrm eq,g})
\left[\omega'_g \left(\frac{\partial P}
{\partial \varepsilon}\right)
\nabla_{k}u^{k}
+p'^{k} \left(\frac{\nabla_{k}P}
{\varepsilon +P}\right. \right. \nonumber\\
&& \left. \left. -\frac{\nabla_{k}T}{T}\right)
 +\frac{p'^{k}p'^{l}}{\omega'_g}
\nabla_{k}u_{l}\right]\bigg\}.
\label{EM_delta2}
\ea

 Let us first discuss the general form of the first-order viscous 
tensor, $\pi_{ij}$ and compare it with the abovementioned form~\eqref{EM_delta2}
derived from the kinetic theory to extract the coefficient of
viscosities. There are two kinds of viscosity, one is shear ($\eta$) 
and other is bulk ($\zeta$). They are  defined as the coefficients in the
viscous stress tensor, $\pi_{ij}$, 
which depends on the 
space derivatives of the velocity. If the velocity 
gradients are small, we may suppose that the momentum transfer
due to the viscosity  depends only on the first 
derivatives of the velocity. To the same approximation,
$\pi_{ij}$ may be supposed a linear function of the 
derivatives $\partial u_i/\partial x_j$. There can be 
no terms in $\pi_{ij}$ independent of
 $\partial u_i/\partial x_j$, since $\pi_{ij}$ must vanish
 for the ${\bf u}=$ constant. Moreover, 
 $\pi_{ij}$ must also vanish when whole fluid is in uniform rotation.
Hence $\pi_{ij}$ must contain the
 symmetrical combinations of the derivatives
 $\partial u_i/\partial x_j$:
 \ba
  \frac{\partial u_i}{\partial x_j}+
  \frac{\partial u_j}{\partial x_i}.
 \ea 
  
 The most general tensor of rank two 
 satisfying the above condition is 
 \ba
 \pi_{ij}=-\eta \bigg(  \frac{\partial v_i}{\partial x_j}+
  \frac{\partial v_j}{\partial x_i}-\frac{2}{3}
 \delta_{ij} \frac{\partial v_l}{\partial x_l}\bigg)
 -\zeta \delta_{ij}  \frac{\partial v_l}{\partial x_l}.
 \label{gen_ten}
 \ea 
The coefficients $\eta$ and $\zeta$ are independent of 
 the velocity, which is true for the isotropic fluid
 because its properties must be described
 by scalar quantities only (in this case, 
 $\eta$ and $\zeta$). The terms in~\eqref{gen_ten} are arranged
 so that the expression in parentheses has the 
 property of vanishing on contraction with respect to 
 $i$ and $j$.

Thus, the traceless part of $\pi^{ij}$ (non-diagonal components)   
is  associated to $\eta$, which measures 
the response of the medium to the fluctuations
in the transverse momentum density and the non-zero trace part
is related to $\zeta$, which is a measure of the response of 
the medium to expansion. That is why, the non-zero trace part 
is related to the diagonal components of $\pi^{ij}$, which
 quantify the change in the pressure of the
medium.  Thus, in the presence 
of  velocity gradients, the leading order 
correction to the energy-momentum
tensor for a rotationally symmetric and isotropic
medium can be parameterized in terms of 
$\eta$ and $\zeta$ as~\cite{Lifshitz:press1981}
\ba
\pi^{ij}=-\eta W_{ij}-\zeta \delta_{ij}\partial_l u^l,
\label{master}
\ea
where
\ba
W_{ij}=\partial_j u^i+\partial_i u^j-\frac{2}{3}
\delta_{ij}\partial_l u^l.
\label{vel_grad2}  
\ea 

Now we replace the velocity gradients by the expression
 \ba
\partial_k u_l=-\frac{1}{2}W_{kl}-
\frac{1}{3}\delta_{kl}\partial_j u^j
\label{vel_grad1}  
\ea
to rewrite the tensor $\pi^{ij}$ obtained 
from the kinetic theory~\eqref{EM_delta2}
into the the generic form~\eqref{master} as
 \begin{eqnarray}
\pi^{ij}&=&2\sum_i g_i\int \frac{d^3p}{(2\pi)^3}
\frac{p^{i}p^{j}}{\omega_i T}\tau_i \bigg\{f_{\mathrm eq,i}
(1-f_{\mathrm eq,i})
\left[\bigg(\omega_i \left(\frac{\partial P}
{\partial \varepsilon}\right)
-\frac{p^2}{3\omega_i}\bigg)\partial_{l}u^{l}
+p^{k} \left(\frac{\nabla_{k}P}{\varepsilon +P}
-\frac{\nabla_{k}T}{T}\right)\right. \nonumber \\
&&\left. -\frac{p^{k}p^{l}}{2\omega_i}W_{kl}\right]
+g_in_{\mathrm eq,i}^{-1} f_{\mathrm eq,i} 
\int_{p'}f_{\mathrm eq,i}(1-f_{\mathrm eq,i})
\left[\bigg(\omega'_i \left(\frac{\partial P}{\partial
 \varepsilon}\right)-\frac{p'^2}{3\omega'_i}\bigg)
\partial_{l}u^{l} \right. \nonumber\\
&& \left. +p'^{k} \left(\frac{\nabla_{k}P}{\varepsilon +P}
-\frac{\nabla_{k}T}{T}\right)
 -\frac{p'^{k}p'^{l}}{2\omega'_i}W_{kl}\right]\bigg\} \nonumber\\
&+& g_g\int \frac{d^3p}{(2\pi)^3}
\frac{p^{i}p^{j}}{\omega_g T}\tau_g
\bigg\{f_{\mathrm eq,g}(1+f_{\mathrm eq,g})
\left[\bigg(\omega_g \left(\frac{\partial P}
{\partial \varepsilon}\right)-\frac{p^2}{3\omega_g}\bigg)
\partial_{l}u^{l} +p^{k} \left(\frac{\nabla_{k}P}{\varepsilon +P}
-\frac{\nabla_{k}T}{T}\right)\right.\nonumber\\
&&\left. -\frac{p^{k}p^{l}}{2\omega_g}W_{kl}\right]
+g_gn_{\mathrm eq,g}^{-1} f_{\mathrm eq,g}
  \int_{p'}f_{\mathrm eq,g}(1+f_{\mathrm eq,g})
\left[\bigg(\omega'_g \left(\frac{\partial P}
{\partial \varepsilon}\right)
-\frac{p'^2}{3\omega'_g}\bigg)
\partial_{l}u^{l} \right. \nonumber \\
&&\left. +p'^{k} \left(\frac{\nabla_{k}P}{\varepsilon +P}
-\frac{\nabla_{k}T}{T}\right)\right.
\left. -\frac{p'^{k}p'^{l}}{2\omega'_g}W_{kl}\right]\bigg\}.
\label{diss_part}
\end{eqnarray} 

The comparison of Eqns.~\eqref{diss_part} and~\eqref{master} thus
yields the shear  and bulk viscosities, which is decomposed into
RT contribution and some correction factor as 
\ba
\eta=\eta^{\rm RT}+\eta^{\rm corr}, 
\ea
where
\ba 
\eta^{\rm RT}&=&\frac{2\beta}{15} 
\sum_i g_i \tau_i\int \frac{d^3p}{(2\pi)^3}
~\frac{p^4}{\omega_i^2}~f_{\mathrm eq,i}
(1-f_{\mathrm eq,i})\nonumber\\
&&+\frac{\beta}{15}  g_g ~\tau_g~\int \frac{d^3p}{(2\pi)^3}
~\frac{p^4}{\omega_g^2}~f_{\mathrm eq,g}
(1+f_{\mathrm eq,g}),\\
\eta^{\rm corr}&=& \frac{2\beta}{15} 
\sum_i g_i^2n_{\mathrm eq,i}^{-1}
 \tau_i~\int \frac{d^3p}{(2\pi)^3}
\frac{p^2}{\omega_i} f_{\mathrm eq,i}(p) 
\int_{p'}\frac{p'^2}{\omega'_i}
f_{\mathrm eq,i}(p')(1-f_{\mathrm eq,i}(p'))\nonumber\\
&&+ \frac{\beta}{15}  g_g^2~n_{\mathrm eq,g}^{-1}
 ~\tau_g~\int \frac{d^3p}{(2\pi)^3}
~\frac{p^2}{\omega_g} ~f_{\mathrm eq,g}(p) 
\int_{p'}\frac{p'^2}{\omega'_g}
~f_{\mathrm eq,g}(p')(1+f_{\mathrm eq,g}(p')).
\ea
Similarly, the bulk viscosity can also be written as 
\ba
\zeta=\zeta^{\rm RT}+\zeta^{\rm corr}, 
\ea
where
\ba \label{bulk_RT}
 \zeta^{\rm RT}&=&\frac{2}{3} \sum_i g_i \int \frac{d^3p}{(2\pi)^3}
\frac{p^2}{\omega_i}~f_{\mathrm eq,i}
(1-f_{\mathrm eq,i})~A_{i,1}\nonumber\\
&&+\frac{1}{3}  g_g \int \frac{d^3p}{(2\pi)^3}
\frac{p^2}{\omega_g}~f_{\mathrm eq,g}(1+f_{\mathrm eq,g})~A_{g,1},\\
\label{bulk_corr}
\zeta^{\rm corr}&=&\frac{2}{3} \sum_i g_i^2~n_{\mathrm eq,i}^{-1}~ 
\int \frac{d^3p}{(2\pi)^3}
~\frac{p^2}{\omega_i} ~f_{\mathrm eq,i}(p) \int_{p'}
f_{\mathrm eq,i}(p')(1-f_{\mathrm eq,i}(p'))~A_{i,2}\nonumber\\
&&+ \frac{1}{3}  g_g^2~n_{\mathrm eq,g}^{-1} ~\int \frac{d^3p}{(2\pi)^3}
\frac{p^2}{\omega_g} f_{\mathrm eq,g} (p) \int_{p'}
~f_{\mathrm eq,g}(p')(1+f_{\mathrm eq,g}(p'))~A_{g,2},
\ea
where 
\ba \label{b1}
A_{i,1}&=&\frac{\tau_i}{3T}\bigg\{\frac{p^2}{\omega_i}
-3\omega_i\left(\frac{\partial P}{\partial \varepsilon}\right)\bigg\},\\
\label{b2}
A_{g,1}&=&\frac{\tau_g}{3T}\bigg\{\frac{p^2}{\omega_g}
-3\omega_g\left(\frac{\partial P}{\partial \varepsilon}\right)\bigg\},\\
\label{b3}
A_{i,2}&=&\frac{\tau_i}{3T}\bigg\{\frac{p'^2}{\omega'_i}
-3\omega'_i\left(\frac{\partial P}{\partial \varepsilon}\right)\bigg \},\\
\label{b4}
A_{g,2}&=&\frac{\tau_g}{3T}\bigg\{\frac{p'^2}{\omega'_g}
-3\omega'_g\left(\frac{\partial P}{\partial \varepsilon}\right)\bigg \}.
\ea 
To calculate the bulk viscosity, 
the Landau-Lifshitz condition,  
 which requires $``00"$ component of $\pi^{\mu\nu}$ 
 to vanish ({\em i.e} $\pi^{00}=0$), should be satisfied 
 in the local rest 
 frame. For that purpose, we have replaced 
   $A_{i,1}$, $A_{i,2}$, $A_{g,1}$ and $A_{g,2}$ by
$A'_{i,1} \rightarrow (A_{i,1}-b_i~\omega_i) $, 
$A'_{g,1} \rightarrow (A_{g,1}-b_g~\omega_g) $,
$A'_{i,2} \rightarrow (A_{i,2}-b'_i~\omega'_i) $, and
$A'_{g,2} \rightarrow (A_{g,2}-b'_g~\omega'_g )$, respectively 
in Eqn~\eqref{EM_delta1} such that
~\cite{Chakraborty:PRC83'2011}
\ba \label{c1}
 \sum_i g_i \int \frac{d^3p}{(2\pi)^3}
~\omega_i~f_{\mathrm eq,i}(1-f_{\mathrm eq,i})~
(A_{i,1}-b_i~\omega_i)=0,\\
\label{c2}
 \sum_i g_i^2~n_{\mathrm eq,i}^{-1}~ 
\int \frac{d^3p}{(2\pi)^3}
~\omega_i ~f_{\mathrm eq,i}(p) \int_{p'}
f_{\mathrm eq,i}(p')(1-f_{\mathrm eq,i}(p'))~
(A_{i,2}-b_i'~\omega'_i)=0,\\
\label{c3}
   \int \frac{d^3p}{(2\pi)^3}
\omega_g~f_{\mathrm eq,g}(1+f_{\mathrm eq,g})~
(A_{g,1}-b_g~\omega_g)=0,\\
\label{c4}
  ~\int \frac{d^3p}{(2\pi)^3}
\omega_g~f_{\mathrm eq,g}(p) \int_{p'}
~f_{\mathrm eq,g}(p')(1+f_{\mathrm eq,g}(p'))~
(A_{g,2}-b_g'~\omega'_g) = 0.
\ea 
We have evaluated the constants $b_i, b_g, b_i'$ and $b_g'$
from the Eqns.~\eqref{c1},~\eqref{c2},~\eqref{c3} and \eqref{c4}, 
respectively and have replaced 
$A_{i,1}$, $A_{i,2}$, $A_{g,1}$ and $A_{g,2}$ by 
$A'_{i,1}$, $A'_{i,2}$, $A'_{g,1}$ and $A'_{g,2}$, 
respectively in Eqns.
\eqref{bulk_RT} and \eqref{bulk_corr} to get final expressions 
of the bulk viscosity   
 \ba
\zeta^{\rm RT}&=&\frac{2\beta}{9} \sum_i g_i\tau_i \int \frac{d^3p}{(2\pi)^3}
~f_{\mathrm eq,i}(1-f_{\mathrm eq,i})
\left \{\frac{p^2}{\omega_i}-3\omega_i \left(\frac{\partial P}
{\partial \varepsilon}\right)\right \}^2\nonumber\\
&&+\frac{\beta}{9}  g_g \tau_g \int \frac{d^3p}{(2\pi)^3}
 ~ f_{\mathrm eq,g}(1+f_{\mathrm eq,g})
\bigg \{\frac{p^2}{\omega_g}
-3\omega_g\left (\frac{\partial P}
{\partial \varepsilon}\right)\bigg \}^2,\\
\zeta^{\rm corr}&=&\frac{2\beta}{9} 
\sum_i g_i^2~n_{\mathrm eq,i}^{-1}\tau_i~ 
\int \frac{d^3p}{(2\pi)^3}
~\frac{p^2}{\omega_i} ~f_{\mathrm eq,i} \int_{p'}
f_{\mathrm eq,i}(p)(1-f_{\mathrm eq,i}(p))
 \bigg \{\frac{p'^2}{\omega'_i}
-3 \omega'_i\left(\frac{\partial P}{\partial \varepsilon}
\right)-b'_{i}~\omega'_i\bigg \}\nonumber\\
&+&\frac{\beta}{9}  g_g^2~n_{\mathrm eq,g}^{-1}
 \tau_g~\int \frac{d^3p}{(2\pi)^3}
\frac{p^2}{\omega_g} f_{\mathrm eq,g}(p) 
 \int_{p'}
~f_{\mathrm eq,g}(p')(1+f_{\mathrm eq,g}(p'))
\bigg \{\frac{p'^2}{\omega'_g}
-3\omega'_g\left(\frac{\partial P}{\partial \varepsilon}\right)
-b'_{g}~ \omega'_g\bigg \},\nonumber\\
\ea
where the constants $b'_{i}$ and $b'_{g}$ are given by
\ba
 b'_{i}&=&\frac{ \sum_i g_i^2~n_{\mathrm eq,i}^{-1}~ 
\int \frac{d^3p}{(2\pi)^3}
~\omega_i ~f_{\mathrm eq,i}(p) \int_{p'}
f_{\mathrm eq,i}(p')(1-f_{\mathrm eq,i}(p'))
\left \{ \frac{p'^2}{\omega'_i}
-3\omega'_i \left(\frac{\partial P}
{\partial \varepsilon}\right)\right\}}
{ \sum_i g_i^2~n_{\mathrm eq,i}^{-1}~ 
\int \frac{d^3p}{(2\pi)^3}
~\omega_i ~f_{\mathrm eq,i}(p) \int_{p'}
f_{\mathrm eq,i}(p')(1-f_{\mathrm eq,i}(p'))\omega'_i},\\
b'_{g}&=&\frac{\int \frac{d^3p}{(2\pi)^3}
\omega_g f_{\mathrm eq,g}(p) \int_{p'}
~f_{\mathrm eq,g}(p')(1+f_{\mathrm eq,g}(p'))
~\left\{\frac{p'^2}{\omega'_g}
-3\omega'_g \left(\frac{\partial P}
{\partial \varepsilon}\right)\right \}}
{\int \frac{d^3p}{(2\pi)^3}
\omega_g f_{\mathrm eq,g}(p)\int_{p'}
~f_{\mathrm eq,g}(p')(1+f_{\mathrm eq,g}(p'))~\omega'_g}.
\ea

 Now we could visualize the correspondence between 
the collision integrals and the momentum transport 
coefficients by computing
the shear ($\eta$) and bulk ($\zeta$) viscosities as a function of 
temperature (seen in Fig.~\ref{bulk}).
We found that $\eta$ for the medium with BGK collsion term is always larger
than ($\sim$ 1.7 times) its value with the simplified
RT term whereas $\zeta$ in BGK becomes smaller ($\sim$ 0.7 times) than the
RT value. This happens due to the opposite behaviours manifested by the 
correction factors, $\eta^{\rm corr}$ and $\zeta^{\rm corr}$,
{\em i.e.} $\eta^{\rm corr}$ is always positive and relatively large while
$\zeta^{\rm corr}$ is always negative and relatively small. This implies
that BGK collison term, {\em especially} the extra 
term in~\eqref{bgk_col} afffect 
shear and bulk viscosities differently.
\begin{center}
\begin{figure}[H]
\begin{tabular}{cc}
\includegraphics[width=6cm]{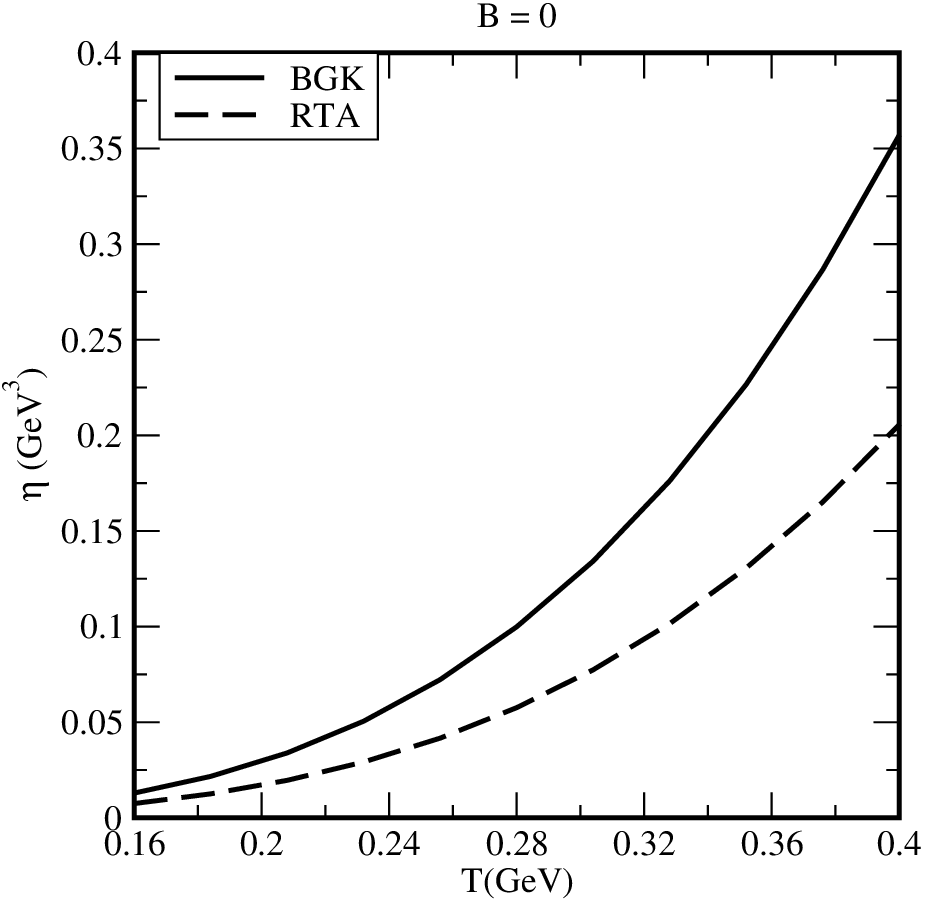}&
\hspace{2.5cm}
\includegraphics[width=6cm]{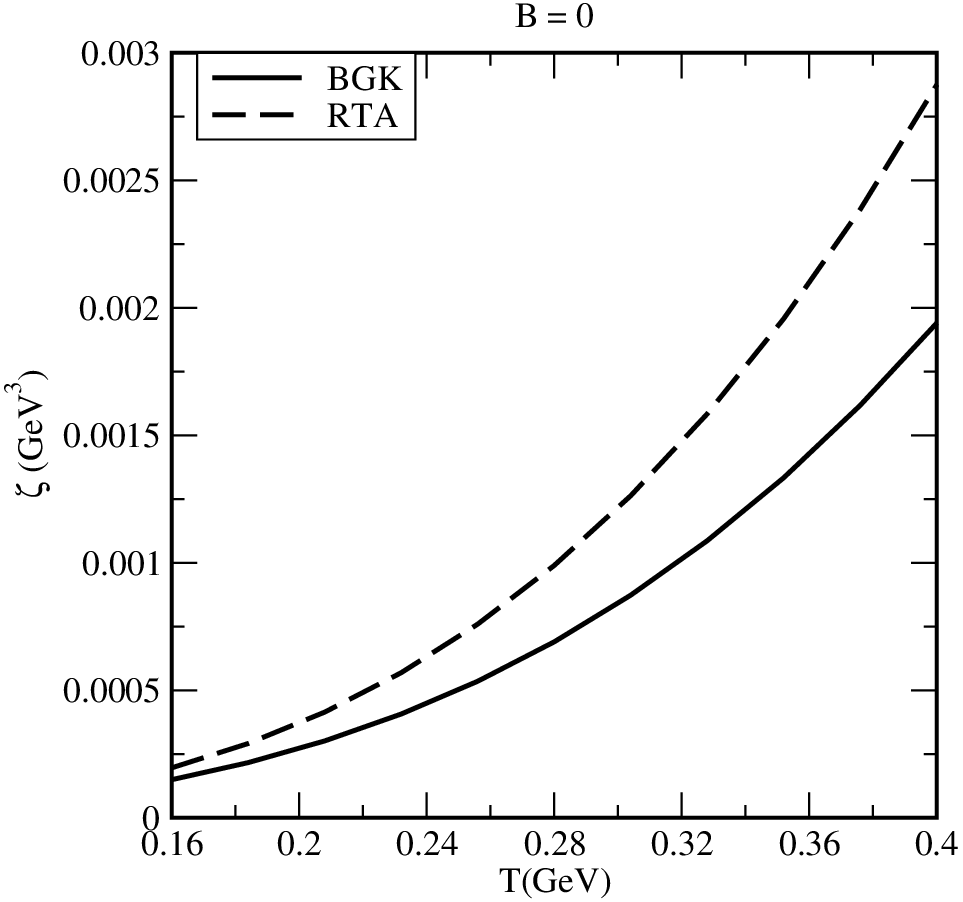}\\
(a)& \hspace{2.5cm}(b)
\end{tabular}
\caption{Shear viscosity (a) and bulk viscosity (b)
 as a function of $T$ in the absence of 
strong magnetic field} 
\label{bulk} 
\end{figure}
\end{center}

Having understood the relation between the transport coefficients: $\eta$ 
and $\zeta$ and the input in the transport equation: BGK and RT collision
terms, we wish to see how a external strong magnetic field could affect
the aforesaid relation in the next subsection.
\subsection{Shear $(\eta_B)$ and bulk  $(\zeta_B)$ viscosities in 
 the presence of strong magnetic field}\label{3.2}
 In this section, we will evaluate  
 $\eta$ and $\zeta$ viscous coefficients
 of the  magnetized hot QCD medium. The presence of
   magnetic field leads to  
 the quantization of the quark energy in terms of the 
Landau levels as~\cite{Gusynin:NPB462'1996}
\ba
\omega_i=\sqrt{p_3^2+m_i^2+2n|q_iB|},  
\ea
where $n=0,1,2....$ corresponds to various Landau levels. In addition to
this, the phase space integral also gets modified as~\cite{Gusynin:NPB462'1996}
\ba
 \int \frac{d^3p}{(2\pi)^3} \rightarrow 
 \sum_{n=0}^{\infty}\frac{|q_iB|}{2\pi}\int \frac{dp_3}{2\pi} (2-\delta_{n0}).
\ea 
In the SMF limit ($|q_iB|>> T^2$), there is a huge energy gap 
between the LLL and the higher landau levels (HLLs), hence,
only  $n=0$ state is populated. In this situation, motion of 
the quarks gets restricted in the transverse direction and 
becomes purely longitudinal (along the direction 
of the magnetic field {\em i.e. $\mathbf{B}=B\hat{z}$}). The quark contribution 
to the energy-momentum tensor will get modified (the gluon
part remains as it is because gluons are not 
affected by the magnetic field directly\footnote{ 
Gluon thermal mass will get magnetic field dependence 
due to the modification of the quark loop contribution
to the gluon self-energy}). The quark part
in LLL  takes the form
\ba
{T}_{B,q}^{\mu\nu} =2\sum_i \frac{g_i|q_iB|}{4\pi^2}\int dp_3~
\frac{p^{\mu}p^{\nu}}{\omega_i}~f_{i}^B,
\ea
where $f_{i}^B=f_{\rm eq,i}^B+\delta f_{i}^B$. 
$f_{\rm eq,i}^B$ is the equilibrium distribution 
function 
\begin{eqnarray}
f_{\mathrm eq,i}^{B}=\frac{1}{e^{\beta \omega_i}+1},
\end{eqnarray}
and $\omega_i=\sqrt{p_3^2+m_i^2}$. Now, we assume
that the system has slightly  deviated  
from the equilibrium, the dissipative part of the 
 energy-momentum tensor reads as
\ba
 \pi_{B,q}^{\mu\nu} =\sum_i \frac{g_i|q_iB|}{2\pi^2}\int dp_3~
\frac{p^{\mu}p^{\nu}}{\omega_i}~\delta f_{i}^B,
\label{EMT_B}
\ea
where $\delta f_{i}^B$ is the deviation in the 
distribution function of the quarks.
 In the effective (1+1) dimensional kinetic 
theory in the SMF, the RBTE for quarks  takes the form
\ba
p^{\mu}\partial_{\mu}f^B_i(x,p)&=&-p^{\mu}u_{\mu}\nu_i^B 
\left(f^B_i-n^B_{i}n{^B}_{\mathrm eq,i}^{-1} f^B_{\mathrm eq,i}\right).
\label{RBTE_B}
\ea
Here $p^{\mu} \equiv (p^0,0,0,p^{3})$, $x^{\mu} \equiv (x^0,0,0,x^{3})$   
and $n_{\mathrm eq,i}^{B}$ refers 
to the equilibrium number 
density of the quarks which reads  
\ba
n_{\mathrm eq,i}^{B}&=&\frac{g_i|q_iB|}{4\pi^2}
\int dp_3 ~f_{\rm eq,i}^B,
\ea
and $\nu_i^B$ is collision frequency which is  
given by the inverse of the relaxation time. 
In the strong $B$, $\tau_i^B$
 depends on the longitudinal component of the momentum $p_3$
 {\em unlike}  
in pure thermal medium \eqref{tau_B0} where $\tau_i$ 
remains constant and does not depend on momentum. 
The relaxation time 
 has been computed in the presence of strong 
 $B$~\cite{Hattori:PRD95'2017}
\begin{eqnarray}
\tau_i^B(p_3;T,|q_iB|) = \frac{\omega_i (e^{\beta\omega_i}-1)}
{\alpha_s C_Fm_i^2(e^{\beta\omega_i}+1)}
{\left(\int \frac{dp'_3}{\omega'_ i
(e^{\beta \omega'_i}+1)}\right)}^{-1},
\end{eqnarray}
where $C_F$ (=4/3) is the Casimir factor and $\alpha_s$ is
the QCD coupling in the strong magnetic field~\eqref{coupling_B}.

We can solve the RBTE~\eqref{RBTE_B}
 upto first order to get the $\delta f^B_{i}$ as
 \footnote{We use the symbol $\int_{p'_3}$
  for  momentum
integration in strong $B$, 
$\int_{p'_3} =\frac{|q_iB|}{2\pi}\int \frac{{dp'_3}}{2\pi} $}.
\ba
\delta f^B_{i}&=&\delta f{^B}_i^{(0)}
+ g_in_{\mathrm eq,i}^{-1} f_{\mathrm eq,i} 
\int_{p_3'}\delta f{^B}_i^{(0)},
\ea
where
\ba
 \delta f{^B}_i^{(0)}&=&-\frac{\tau_i^B}{p^{\nu}u_{\nu}} ~
  p^{\mu}\partial_{\mu}f^B_i(x,p)\nonumber\\
&=&-\frac{\tau_i^B}{\omega_i T} ~
  p^{\mu}f^B_{\mathrm eq,i}(1-f^B_{\mathrm eq,i})
\left[u_{\alpha}p^{\alpha}u_{\mu}\frac{DT}{T}+
u_{\alpha}p^{\alpha}\frac{\nabla_{\mu}T}{T}-u_{\mu}p^{\alpha}Du_{\alpha}
-p^{\alpha}\nabla_{\mu}u_{\alpha}\right].
\ea

Substituting $\delta f^B_{i}$ in~\eqref{EMT_B}, we finally obtain
the spatial component of stress-energy tensor in the strong $\vec{B}$ as
\ba
\pi_{B,q}^{ij}&=&\sum_i \frac{g_i|q_iB|}{2\pi^2}\int dp_3~
\frac{p^{i}p^{j}}{\omega_i T}\bigg\{\tau^B_i f{^B}_{\mathrm eq,i}
(1-f^{B}_{\mathrm eq,i})
\left[\bigg(\omega_i \left(\frac{\partial P}
{\partial \varepsilon}\right)
-\frac{p_3^2}{3\omega_i}\bigg)
\partial_{l}u^{l}
+p^{k} \left(\frac{\nabla_{k}P}{\varepsilon +P}
-\frac{\nabla_{k}T}{T}\right)\right. \nonumber\\
&&\left. -\frac{p^{k}p^{l}}{2\omega_i}W_{kl}\right]
+g_in{^B}_{\mathrm eq,i}^{-1} f{^B}_{\mathrm eq,i} 
\int_{p_3'}\tau^B_i
f{^B}_{\mathrm eq,i}(1-f{^B}_{\mathrm eq,i})
\left[\bigg(\omega'_i \left(\frac{\partial P}
{\partial \varepsilon}\right)
-\frac{p_3^{'2}}{3\omega'_i}\bigg)
\partial_{l}u^{l}\right.\nonumber\\
&&\left. +p'^{k} \left(\frac{\nabla_{k}P}{\varepsilon +P}
-\frac{\nabla_{k}T}{T}\right)\right.
\left. -\frac{p'^{k}p'^{l}}{2\omega'_i}W_{kl}\right]\bigg\}.
\label{disspart_B}
\ea

 Let us now understand the generic form of 
the tensorial structure of 
$\pi_{ij}$ in an external $\vec{B}$. 
The number of independent
tensorial combinations (coefficients of viscosity)
 gets increased from two
(in the absence of $\vec{B}$)
to eight in the presence of $\vec{B}$. However, the number becomes 
seven due to the Onsager relation, out of which, 
five are the coefficients of shear-term, one is bulk-term 
and the last one is due to cross term between 
 the ordinary and volume viscosities. 
 In a medium (QGP), the cross term
vanishes, moreover, in the presence of 
strong $\vec{B}$, the non-diagonal 
terms will also be absent due to 
the vanishing of transverse 
components of the velocity (artifact of strong $\vec{B}$).
Thus, the tensorial structure gets 
reduced in a much simpler form,  leaving
the longitudinal components of the tensor survived, as 
\begin{eqnarray}
&&\pi_{xx} = -\eta_0\left(V_{zz}-\frac{1}{3}\nabla\cdot
\mathbf{V}\right)+\zeta_0\nabla\cdot\mathbf{V} , \\
&&\pi_{yy} = -\eta_0\left(V_{zz}-\frac{1}{3}\nabla\cdot
\mathbf{V}\right)+\zeta_0\nabla\cdot\mathbf{V}, \\
&&\pi_{zz} = 2\eta_0\left(V_{zz}-\frac{1}{3}\nabla\cdot
\mathbf{V}\right)+\zeta_0 \nabla\cdot\mathbf{V}
,\end{eqnarray}
where $\eta_0$ and $\zeta_0$ are known as the longitudinal 
viscosities and the other symbols are given 
in~\cite{Rath:PRD102'2020}.
\footnote{The term longitudinal signifies the direction 
of the velocity with respect to the direction of magnetic field.} 

Similar to the case in the absence of magnetic 
field~\eqref{master}, the 
above components are grouped into the traceless and nonzero trace 
parts as
\begin{eqnarray}
&&\pi_{xx}=\pi_{yy}=-\frac{1}{2}\pi_{zz}=-\eta_0
\left(V_{zz}-\frac{1}{3} {\nabla\cdot \mathbf{V}|}_z \right), \\ 
&&\pi_{xx}=\pi_{yy}=\pi_{zz}=\zeta_0 {\nabla\cdot\mathbf{V}|}_z
,\end{eqnarray}
respectively. Therefore, the spatial component of the viscous
tensor in strong $\vec{B}$ can be expressed in a form 
(by relabeling $\eta_0 \equiv \eta_B$ and $\zeta_0 \equiv \zeta_B$)
\ba
\pi_{B,q}^{ij}=-\eta_B W_{ij}-\zeta_B \delta_{ij}\partial_l u^l.
\label{formula_B}
\ea

Now we can extract the coefficients $\eta_B$ and $\zeta_B$ from
the coefficients of $W_{ij}$ and $\partial_l u^l$ in $\pi^{ij}$ 
obtained
from kinetic theory~\eqref{disspart_B}.
Since gluons are not directly affected by the 
strong magnetic field, we will take the gluon 
contribution~\eqref{diss_part} from
Section 3.1 (in the absence of the magnetic field). 
Finally, the shear  
viscosity in the BGK collision term  is expressed in terms of RT contribution
as
\ba
\eta_{B}= \eta_B^{\rm RT} +\eta_B^{\rm corr},
\ea
 where
\ba
\eta_B^{\rm RT}&=&\frac{\beta}{4\pi^2} \sum_i g_i|q_iB|\int dp_3~ 
\frac{p_3^4}{\omega_i^2}\tau^B_i f{^B}_{\mathrm eq,i}
(1-f^{B}_{\mathrm eq,i})+\frac{\beta}{15}  g_g \tau_g\int \frac{d^3p}{(2\pi)^3}
\frac{p^4}{\omega_g^2}~f_{\mathrm eq,g}(1+f_{\mathrm eq,g})\\
\eta_B^{\rm corr}&=&\frac{\beta}{4\pi^2} \sum_i g_i^2|q_iB|
n{^B}_{\mathrm eq,i}^{-1}\int dp_3~ 
\frac{p_3^2}{\omega_i(p_3)} f{^B}_{\mathrm eq,i}(p_3) 
\int_{p_3'}\frac{{p'_3}^2}{\omega_i(p'_3)}
\tau^B_i(p'_3)f{^B}_{\mathrm eq,i}(p'_3)(1-f{^B}_{\mathrm eq,i}(p'_3))\nonumber\\
&&+\frac{\beta}{15}  g_g^2~n_{\mathrm eq,g}^{-1} ~\tau_g\int \frac{d^3p}{(2\pi)^3}
\frac{p^2}{\omega_g(p)}~ f_{\mathrm eq,g}(p) \int_{p'}\frac{p'^2}{\omega_g(p')}
f_{\mathrm eq,g}(p')(1+f_{\mathrm eq,g}(p')).
\ea 

 We will now depict how the strong $B$ could 
 modulate the correspondence
between the collision integrals and the 
(momentum) transport coefficients in Fig.~\ref{shear}. For a thermal 
medium in a strong $B$ environment, the
dominant scale will now be the magnetic field ($|q_fB| >T^2$), unlike the 
temperature is the dominant scale in a thermal medium in the absence of 
magnetic field ($B=0$).
\begin{center}
\begin{figure}[H]
\begin{tabular}{cc}
\includegraphics[width=6cm]{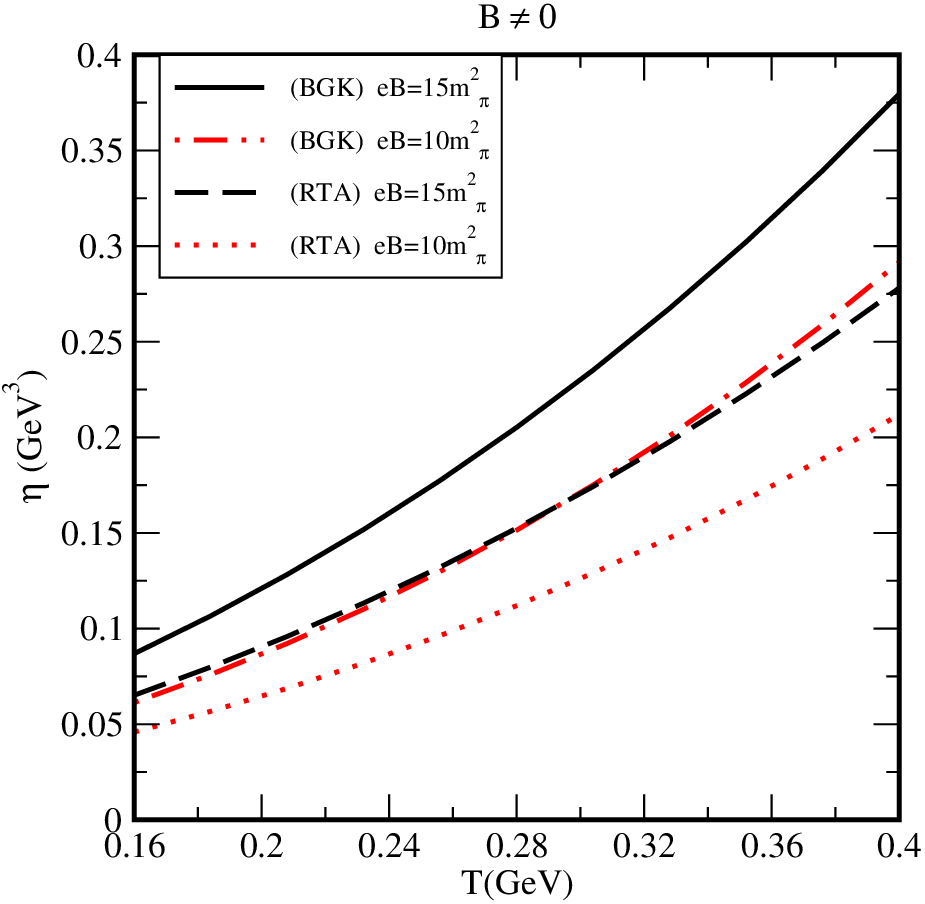}&
\hspace{2.5cm}
\includegraphics[width=6cm]{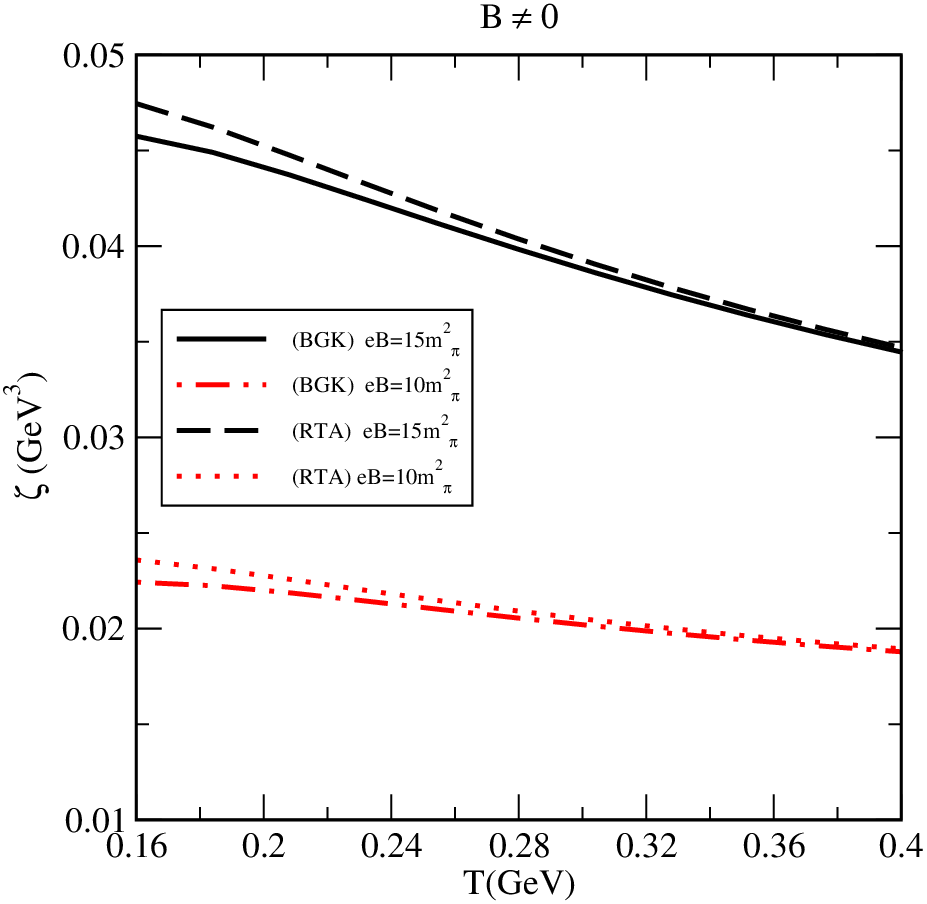}\\
(a)& \hspace{2.5cm} (b)
\end{tabular}
\caption{ Shear viscosity (a) and bulk 
viscosity (b) as a function of $T$ in 
   the presence of strong $B$. } 
\label{shear} 
\end{figure}
\end{center}

 We observe that similar to $B=0$ case, 
the shear viscosity in a
strong $B$ with BGK collision term is still 
larger ($\sim$ 1.3 times) than
the simple minded RT term, which is again 
evidenced by the fact that the 
correction term $\eta_B^{\rm corr}$ is always
 positive. In addition, in the presence of
strong $B$, $\eta$ in both collision terms gets enhanced. To be precise, 
the enhancement with RT term is relatively larger than the BGK term.
In strong $B$, the shear viscosity gets enhanced in BGK collision
integral in comparison to RTA at a fixed $B$.
$\eta$ also increases with the strength of the magnetic
field in BGK as well as RTA collision terms.  

 One important observation we notice
 in Figure~\ref{shear} (a)
 is that $\eta$ with 
BGK collision term at lower (strong) $B$, {\em say} 
$eB=10m_{\pi}^2$ 
looks similar to 
$\eta$ with RT term at higher (strong) $B$, {\em say}
 $eB=15m_{\pi}^2$. 
This leads to an ambiguity in the
phenomenological  modeling while extracting the physical parameters 
of the system by comparing the results from theory to 
the experiments. It means that if one were to extract
the physical parameters from a comparison of theory to experiment one would deduce different values for the
strength of B whether on uses BGK or RTA.
In this case, it would be better to take  results from
BGK collision integral to compute the physical parameters
since it shows an improvement over the naive RTA in the 
sense that it conserve the particle number and charge
 instantaneously.

Similarly, the bulk viscosity can be decomposed as
\ba
\zeta_B=\zeta^{\rm RT}_B+\zeta^{\rm corr}_B,
\label{zeta_B}
\ea
where 
\ba
\zeta_B^{\rm RT}&=& \frac{1}{2\pi^2}\sum_i g_i|q_iB|\int dp_3~ 
 \frac{p_3^2}{\omega_i} ~f{^B}_{\mathrm eq,i}
(1-f^{B}_{\mathrm eq,i})~A^B_{i,1} \nonumber\\
&&+\frac{1}{3}  g_g \int \frac{d^3p}{(2\pi)^3}
\frac{p^2}{\omega_g}~f_{\mathrm eq,g}(1+f_{\mathrm eq,g})~A_{g,1},\\
\zeta_B^{\rm corr}&=& \frac{1}{2\pi^2}\sum_i g^2_i|q_iB|
n{^B}_{\mathrm eq,i}^{-1}\int dp_3~ 
 \frac{p_3^2}{\omega_i} f{^B}_{\mathrm eq,i}(p_3) \int_{p_3'}
f{^B}_{\mathrm eq,i}(p'_3)(1-f{^B}_{\mathrm eq,i}(p'_3))~A^B_{i,2}\nonumber\\
&&+ \frac{1}{3}  g_g^2~n_{\mathrm eq,g}^{-1} ~\int \frac{d^3p}{(2\pi)^3}
\frac{p^2}{\omega_g} f_{\mathrm eq,g}(p) \int_{p'}
f_{\mathrm eq,g}(p')(1+f_{\mathrm eq,g}(p'))~A_{g,2},
\ea
where the factors $A^B_{i,1}$ and $A^B_{i,2}$  are 
\ba 
A^B_{i,1}&=&\frac{\tau^B_i}{3T}\bigg \{\frac{p_3^2}{\omega_i}
-3~\omega_i\left(\frac{\partial P}{\partial \varepsilon}\right)\bigg \},\\
A^B_{i,2}&=&\frac{\tau^B_i}{3T}\bigg \{\frac{p'_3{^2}}{\omega_i}
-3~\omega_i\left(\frac{\partial P}{\partial \varepsilon}\right)\bigg \}.
\ea
 Applying the Landau-Lifshitz condition
 ({\em i.e.}~$\pi_{B,q}^{00}=0$) and following
the similar steps as the $B=0$ case, we get the 
final expression for the  bulk viscosity of the strongly magnetized
thermal QCD medium~\eqref{zeta_B} as
\ba
\zeta_B^{\rm RT}&=& \frac{\beta}{6\pi^2}\sum_i g_i|q_iB|\int dp_3~ 
 \tau^B_i ~f{^B}_{\mathrm eq,i}
(1-f^{B}_{\mathrm eq,i})~\bigg \{\frac{p_3^2}{\omega_i}
-3\omega_i\left(\frac{\partial P}{\partial \varepsilon}\right)\bigg \}^2 \nonumber\\
&&+\frac{\beta}{9}  g_g \tau_g \int \frac{d^3p}{(2\pi)^3}
~f_{\mathrm eq,g}(1+f_{\mathrm eq,g})
\bigg \{\frac{p^2}{\omega_g}
-3\omega_g\left(\frac{\partial P}
{\partial \varepsilon}\right)\bigg \}^2,\\
\zeta_B^{\rm corr}&=& \frac{\beta}{6\pi^2}\sum_i g^2_i|q_iB|
n{^B}_{\mathrm eq,i}^{-1}\int dp_3~ 
 \frac{p_3^2}{\omega_i} f{^B}_{\mathrm eq,i}(p_3) \int_{p_3'}\tau^B_i(p'_3)
f{^B}_{\mathrm eq,i}(p'_3)(1-f{^B}_{\mathrm eq,i}(p'_3))\nonumber \\
&& \times \bigg \{\frac{p'_3{^2}}{\omega'_i}
-3\omega'_i\left(\frac{\partial P}{\partial \varepsilon}\right)
-b'_i\omega_i'\bigg \}+ \frac{\beta}{9}  g_g^2~n_{\mathrm eq,g}^{-1}
\tau_g ~\int \frac{d^3p}{(2\pi)^3}
\frac{p^2}{\omega_g} f_{\mathrm eq,g}(p) \int_{p'}
f_{\mathrm eq,g}(p')\nonumber\\
&& \times (1+f_{\mathrm eq,g}(p'))
\bigg \{\frac{p'^2}{\omega'_g}
-3\omega'_g\left(\frac{\partial P}{\partial \varepsilon}\right)
-b'_{g} \omega'_g\bigg \},
\ea
where 
\ba
b'_i= \frac{\sum_i g^2_i|q_iB|
n{^B}_{\mathrm eq,i}^{-1}\int dp_3~ 
\omega_i f{^B}_{\mathrm eq,i}(p_3) \int_{p_3'}
f{^B}_{\mathrm eq,i}(p'_3)(1-f{^B}_{\mathrm eq,i}(p'_3))~
\bigg \{\frac{p'_3{^2}}{\omega'_i}
-3\omega'_i\left(\frac{\partial P}{\partial \varepsilon}\right)\bigg \}}
{\sum_i g^2_i|q_iB|
n{^B}_{\mathrm eq,i}^{-1}\int dp_3~ 
\omega_i f{^B}_{\mathrm eq,i}(p_3) \int_{p_3'}
f{^B}_{\mathrm eq,i}(p'_3)(1-f{^B}_{\mathrm eq,i}(p'_3))\omega'_i}.
\ea

 We have also now 
plotted bulk viscosity ($\zeta$) as a function
of temperature in the presence of strong magnetic field in the
above Fig.~\ref{shear} (right panel). 
It is found that RT collision term is still found 
to dominate over BGK term except at higher temperature, where
their contributions are almost same. This observation is just opposite 
to the observation
in the absence of magnetic field (seen in Fig.~\ref{bulk}), 
where the merger happens
to be in the small temperature region.

\subsection{Charge and heat transport coefficients in thermal
QCD medium} \label{3.3}
In this subsection, we will revisit our earlier 
work~\cite{Khan:PRD104'2021} wherein we have calculated 
the electrical ($\sigma_{el}$) and thermal ($\kappa$) 
conductivities of the 
hot QCD medium in both BGK and RT collision integrals. We will 
use $\sigma_{el}$ and $\kappa$ to study
the relative competition between the various transport 
coefficients in section~\ref{4}. 
\subsubsection{Electrical conductivity}\label{3.3.1}
The electrical conductivity $\sigma_{el}$, which manifests
the ease of electric current flow in the medium
 has been calculated in the kinetic theory framework using the BGK type 
collision integral. It can be decomposed into two parts 
in a similar fashion like  $\eta$ and $\zeta$ 
in the absence of the magnetic field as  
 \ba
 \sigma_{el}=\sigma_{el}^{\rm RT}+\sigma_{el}^{\rm corr}, 
 \ea 
 where 
\ba
\sigma_{\rm el}^{\rm RT} &=& \frac{2\beta}{3\pi^2} 
\sum_i q_i^2 g_i \tau_i 
\int dp ~\frac{p^4}{\omega_i^2} 
 f_{\mathrm eq,i}(p)(1-f_{\mathrm eq,i}(p)),\\
\sigma_{\rm el}^{\rm Corr} &=& \frac{2\beta}{\pi^2} 
\sum_i q_i^2 g^2_i \tau_i n_{\mathrm eq,i}^{-1} 
\int dp~\frac{p^3}{\omega_i} 
f_{\mathrm eq,i}(p)\int_{p^\prime} 
\frac{ p'}{\omega'_i} f_{\mathrm eq,i}(p') (1 - 
f_{\mathrm eq,i}(p')),
\label{sigmaRT}
\ea
and in  the strong magnetic field  
\ba
\sigma_{el}^B=\sigma_{el}^{\rm B,RT}+\sigma_{el}^{\rm B, corr} ,
\ea 
where
\ba
\sigma^{B,\rm RT}_{el} &=& \frac{\beta}{\pi^2} \sum_i q_i^2 g_i
 |q_iB|\int {dp_3}\frac{p_3^2 }{\omega_i^2 } \tau_i^{B}
 f_{\mathrm eq,i}^{B}(1-f_{\mathrm eq,i}^{B}),\\
\sigma^{B,\rm Corr}_{el} &=& \frac{\beta}{\pi^2} \sum_i q_i^2 g^2_i |q_iB|
{n_{\mathrm eq,i}^B}^{-1} \left[ \int {dp_3}\frac{p_3}{\omega_i} f_{\rm eq,i}^B(p_3)
\int_{p'_{3}}\frac{p'_3}{\omega'_i}~\tau_i^B (p_3^\prime)~f_{\rm eq,i}^{B}(p'_3)~
(1-f_{\rm eq,i}^{B}(p'_3))\right].\nonumber\\
\label{sigma_corr} 
\ea 
\subsubsection{Thermal conductivity}
We have also calculated the heat transport coefficient {\em namely}
thermal conductivity ($\kappa$) from the difference between the energy diffusion 
and the enthalpy diffusion using the BGK collision integral. 
It can be written in the absence of 
the magnetic field as
\ba
\kappa =\kappa^{\rm RT}+\kappa^{\rm Corr},
\ea
where 
\ba  
\kappa^{\rm RT} &=& \frac{\beta^2}{3\pi^2} \sum_i g_i \tau_i \int dp
\frac{p^4}{\omega_i^2(p)}(\omega_i(p)-h_i)^2f_{\rm eq,i}(p)
(1-f_{\rm eq,i}(p)),\\
\kappa^{\rm Corr} &=& \frac{\beta^2}{\pi^2} \sum_i g^2_i \tau_i n_{\mathrm eq,i}^{-1}
\int dp \frac{p^3}{\omega_i(p)}(\omega_i(p)-h_i)f_{\rm eq,i}(p)\nonumber\\
&& \times \int_{p^\prime} \frac{p'}{\omega_i(p')}
(\omega_i(p')-h_i)  f_{\rm eq,i}(p')(1-f_{\rm eq,i}(p')),
\label{kappa_corr}
\ea
and in the strong magnetic field as 
\ba 
\kappa^B =\kappa^{\rm B,RT}+\kappa^{\rm B,Corr},
\ea 
where
\begin{eqnarray}
\kappa^{\rm B,RT} &=& \frac{\beta^2}{2\pi^2} \sum_i{g_i|q_iB|}
 \int {dp_3}\frac{p_3^2}
{\omega_i^2 }\tau_i^{B}(\omega_i-h_i^B)^2f_{\rm eq,i}^{B}(1-f_{\rm eq,i}^{B}),
\label{kappa_rt_B},\\
\kappa^{\rm B,Corr} &=& \frac{\beta^2}{2\pi^2} \sum_i{g^2_i|q_iB|}
{n_{\mathrm eq,i}^B}^{-1}  \int {dp_3}\frac{p_3}
{\omega_i}(\omega_i-h_i^B)f_{\rm eq,i}^{B}(p_3)\nonumber\\
&& \times \int_{p'_{3}}\frac{p'_3}
{\omega'_i}\tau_i^{B}(p'_3)
(\omega'_i-h_i^B)f_{\rm eq,i}^{B}(p'_3)(1-f_{\rm eq,i}^{B}(p'_3)).
\label{kappa_B_corr} 
\end{eqnarray}

\section{Applications}\label{4}
 In this section, we will study how  BGK collision integral 
 modifies the specific shear and  bulk viscosities needed 
 to explore the  fluidity and transition point of the 
 QCD phase. We will further check  
  the relative behavior
among the momentum, heat, and charge diffusion in the strongly
magnetized QCD medium. These 
derived coefficients {\em namely} Prandtl number, Reynolds number and 
$\gamma$ factor characterize the various properties {\em like}
degree of sound attenuation in the medium, nature of the flow, etc. At last,
we see the relative competition between the shear and bulk viscosities 
in terms of ratio $\zeta/\eta$.
\subsection{Specific shear ($\eta/s$) and
 bulk  ($\zeta/s$) viscosities }
 We will now focus 
 on  the ratios
 $\eta/s$ and $\zeta/s$, also known as specific 
 shear and specific bulk
 viscosities, respectively. These ratios give an
  idea about the perfectness
and conformal nature of the fluid, respectively. The ratio
$\eta/s$ has been calculated for the QGP using the 
parton transport method~\cite{Xu:PRL101'2008} and was found to be 
very small, confirming the strongly coupled nature of the 
QGP, which nullifies the widespread belief that QGP happens to be 
a weakly interacting 
gas of quarks and gluons. This is also in agreement 
with the famous KSS bound  of the 
AdS/CFT correspondence~\cite{Kovtun:PRL94'2005}. The  
relativistic viscous hydrodynamics 
~\cite{Luzum:PRC78'2008} also uses very  small value of  $\eta/s$ ratio
 (around 0.08 to 0.1)
to reproduce the RHIC data~\cite{Gavin:PRL97'2006}  
 and also  matches well with
lattice calculations~\cite{Nakamura:PRL94'2005}.
The entropy density of the hot QCD medium can be defined using 
the thermodynamical relation 
\ba\label{entropy_den}
s=\frac{(\varepsilon+P)}{T} ,
\ea 
where $\varepsilon$ and $P$ are the energy density
 and pressure, respectively.
First, we calculate  $\varepsilon$ and 
$P$ for $B=0$ case as
\ba \label{energy_b0}
\varepsilon &=&\frac{1}{\pi^2}\sum_{i}g_i
\int dp~ p^2 \omega_i~f_{\mathrm eq,i}+
\frac{g_g}{2\pi^2}\int dp ~p^2~ \omega_g ~f_{\mathrm eq,g},\\
P &=& \frac{1}{{3\pi^2}}\sum_{i}g_i\int dp~ 
\frac{p^4}{\omega_i}~f_{\mathrm eq,i}+
\frac{g_g}{6\pi^2}\int dp ~\frac{p^4}
{\omega_g} ~f_{\mathrm eq,g},
\label{pressure_b0}
\ea
respectively. In the strong magnetic field, we have 
\ba \label{energy_bnot0}
\varepsilon_{B}&=&\frac{1}{2\pi^2}\sum_i g_i|q_iB|\int dp_3~
\omega_i ~f_{\rm eq,i}^{B}+
\frac{g_g}{2\pi^2}\int dp ~p^2~ \omega_g ~f_{\mathrm eq,g},\\
P_{B}&=&\frac{1}{2\pi^2} \sum_i g_i|q_iB|\int dp_3~\frac{p_3^2}
{\omega_i} ~f_{\rm eq,i}^{B}+
\frac{g_g}{6\pi^2}\int dp ~\frac{p^4}{\omega_g} ~f_{\mathrm eq,g}.
\label{pressure_bnot0}
\ea
 The entropy density in  $B=0$ and $B\neq0$ cases 
can be calculated from  Eqn. \eqref{entropy_den} as 
\ba \label{entropy_B0}
s&=&\frac{\beta}{3\pi^2}\sum_{i} g_i\int dp~ p^2
\left(\frac{p^2}{\omega_i}+3\omega_i \right)
f_{\mathrm eq,i}+
\frac{\beta}{6\pi^2}g_g\int dp ~p^2\left(\frac{p^2}{\omega_g}
+3\omega_i\right)f_{\mathrm eq,g},\\
s_{B}&=&\frac{\beta}{2\pi^2} \sum_i g_i|q_iB|\int dp_3~\left(
\frac{p_3^2}
{\omega_i}+\omega_i\right) f_{\rm eq,i}^{B} +
\frac{\beta}{6\pi^2}g_g\int dp ~p^2\left(\frac{p^2}{\omega_g}
+3\omega_i\right)f_{\mathrm eq,g}, 
\label{entropy_B} 
\ea
respectively.
\vspace{5mm}
\begin{center}
\begin{figure}[H]
\begin{tabular}{cc}
\includegraphics[width=6cm]{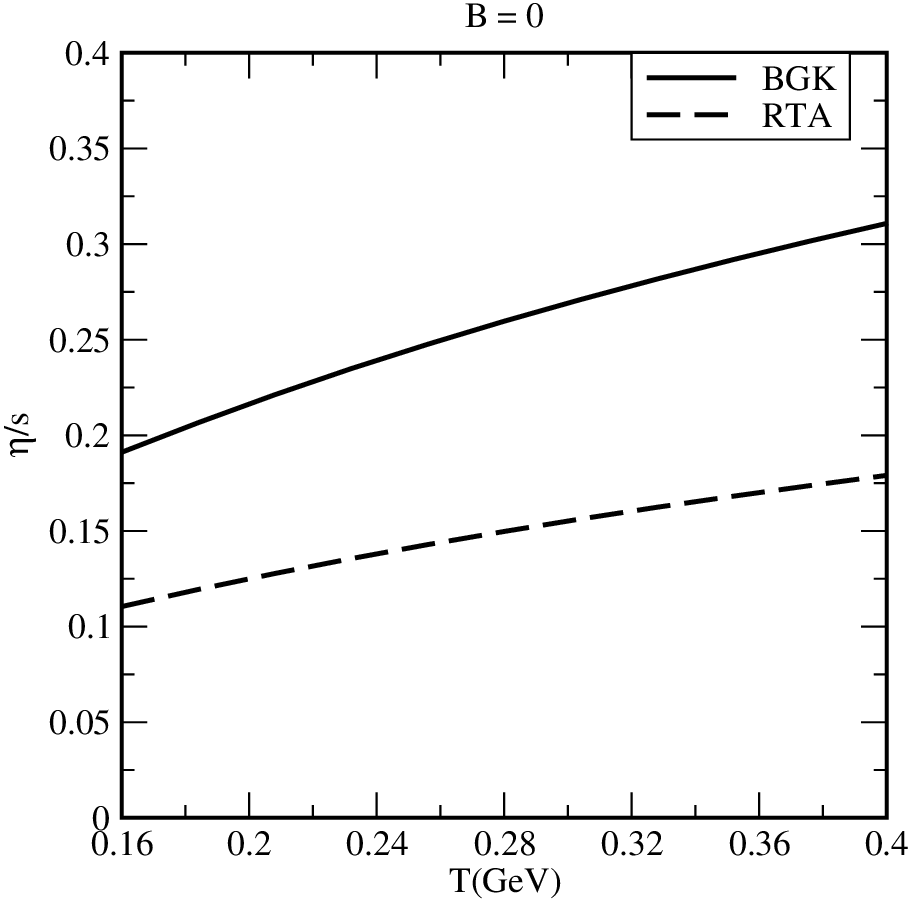}&
\hspace{2.5cm}
\includegraphics[width=6cm]{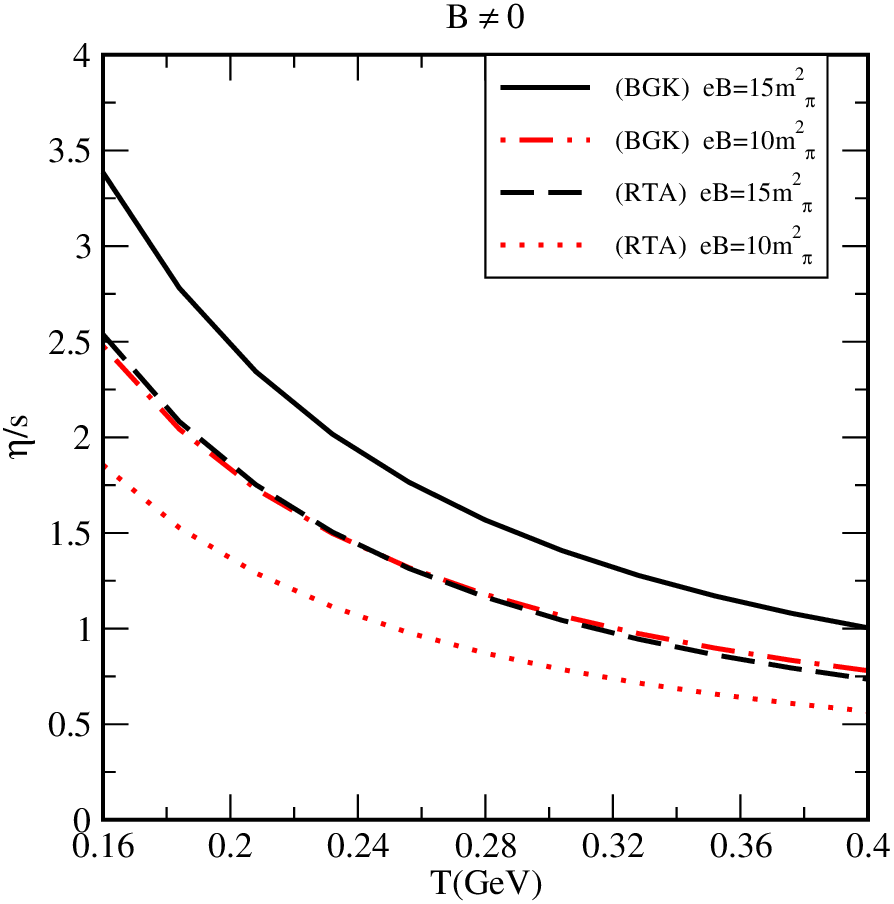}\\
(a) &\hspace{2.5cm} (b)
\end{tabular}
\caption{ Shear viscosity to  entropy density ratio $(\eta/s)$ 
 as a function of $T$ in absence (a)
 and in presence (b) of the strong $B$ } 
\label{shear_s} 
\end{figure}
\end{center}
\begin{center}
\begin{figure}[H]
\begin{tabular}{cc}
\includegraphics[width=6cm]{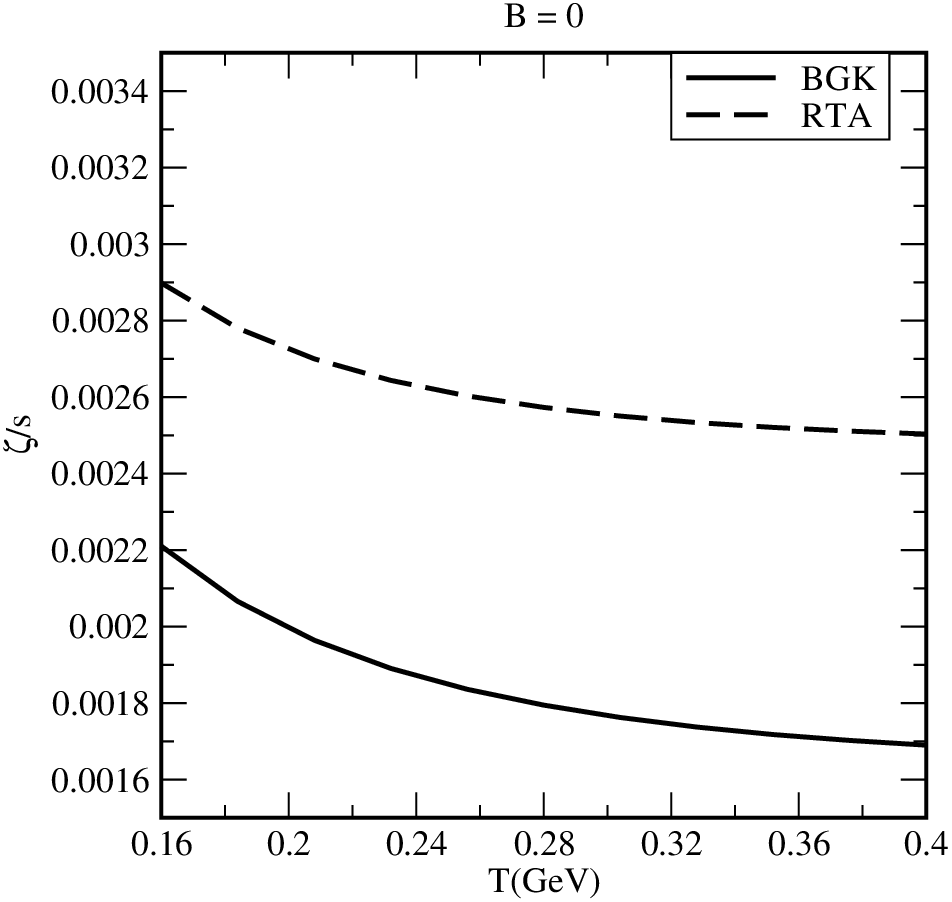}&
\hspace{2.5cm}
\includegraphics[width=6cm]{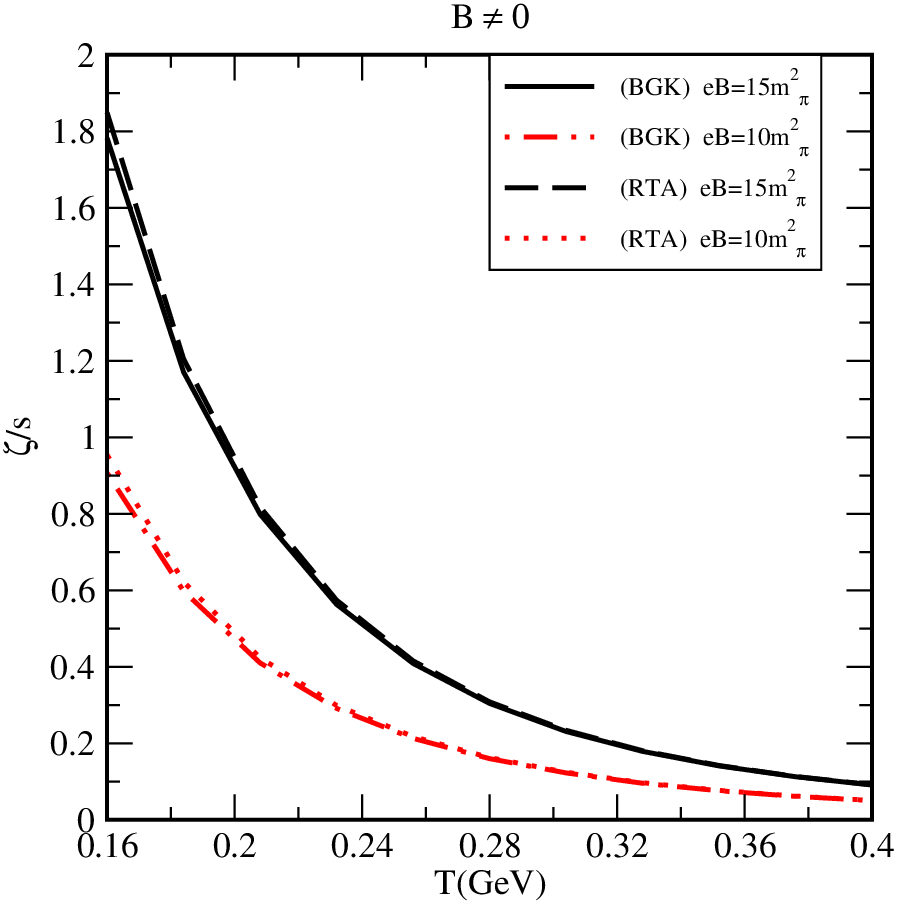}\\
(a)& \hspace{2.5cm}(b)
\end{tabular}
\caption{Bulk viscosity to entropy density ratio $(\zeta/s)$ as a 
function of $T$ in absence (a)
 and in presence (b) of the strong $B$ } 
\label{bulk_s} 
\end{figure}
\end{center}
 
Fig.~\ref{shear_s} shows our estimates of 
 the ratio $\eta/s$ as a function of
temperature in the absence (left panel) and  presence of the strong
$B$ (right panel). The ratio $\eta/s$ increases
 (decreases) with $T$ in the absence (presence) of the 
strong $B$. The magnitude of  $\eta/s$ gets enhanced in the 
BGK collision integral in both cases. It indicates that
instantaneous conservation of the particle number in the 
medium takes the fluid away from its ideal nature. The 
$\eta/s$ ratio gets enhanced in strong $B$. 
 Like the case of $\eta$ in a strong $B$ in 
Fig.~\ref{shear} (a), $\eta/s$ ratio in BGK collision term 
at $eB=10 m_\pi^2$ 
also looks similar to its counterpart with RT term at $eB=15m_\pi^2$.
As we mentioned earlier, this similarity leads to an apparent 
ambiguity in extracting the physical parameters by comparing the theoretical
predictions to the experimental data. It would be
better to prefer BGK collision term over RTA.

In Fig.~\ref{bulk_s}, we have displayed the ratio 
 $\zeta/s$ as a function of
temperature. We found 
that  $\zeta/s$ decreases with 
$T$ in $B=0$ (left panel) as well as in $B\neq0$ 
(right panel) case. The magnitude of
$\zeta/s$ gets reduced in the BGK collision term in comparison
to the RT in the $B=0$ scenario while 
there in the presence of strong $B$, both the 
collision integrals produce
similar results. It further increases as the 
strength of $B$ grows, which may indicate that 
the system moves away from
the conformal nature due to the presence of 
the strong magnetic field.
\subsection{Prandtl number}
  The ratio of the 
momentum diffusivity to the thermal diffusivity in a given medium is 
quantified in terms of the Prandtl number (Pr)
\ba
{\rm Pr}=\frac{\eta/\rho}{\kappa/C_p},
\ea
where $\rho$ is the mass density, $C_P$ denotes 
the specific heat at constant pressure
and $\kappa$ refers to the thermal conductivity of the medium under 
consideration. Pr gives an idea about the roles of the 
shear viscosity and thermal conductivity on the sound attenuation in 
a system. The smaller value of the Pr number (Pr $<<1$) corresponds to  
the dominance of the thermal
diffusion while higher value  (Pr $>>1$) to  
that of momentum diffusion.  
The specific heat $C_P$ can be calculated from the 
thermodynamic relation
\ba
C_p=\frac{\partial}{\partial T}(\varepsilon+P), 
\ea
where $\varepsilon$ and $P$ are the energy density and pressure, 
respectively and have been calculated in the absence [Eqns.~\eqref{energy_b0} and 
\eqref{pressure_b0}] as well as in the presence of 
strong $B$ [Eqns.~\eqref{energy_bnot0} and 
\eqref{pressure_bnot0}]. The specific heat 
$C_P$ can be evaluated  in the absence and 
in presence of the SMF as
\ba
C_p&=&\frac{\beta^2}{3\pi^2}\sum g_i\int dp~p^2
(p^2+3\omega_i^2)~ f_{\rm eq,i}(1-f_{\rm eq,i})\nonumber\\
&&+\frac{\beta^2}{6\pi^2} g_g\int dp~
p^2(p^2+3\omega_g^2)~ f_{\mathrm eq,g}(1+f_{\mathrm eq,g}),
\ea
and
\ba
 C^B_p&=&\frac{\beta^2}{2\pi^2}\sum_i g_i|q_iB|\int dp_3~
(p_3^2+\omega_i^2)~f_{\rm eq,i}^{B}(1-f_{\mathrm eq,i}^{B})\nonumber\\
&&+\frac{\beta^2}{6\pi^2} g_g\int dp~
p^2(p^2+3\omega_g^2)~ f_{\mathrm eq,g}(1+f_{\mathrm eq,g}), 
 \ea
 respectively. Another quantity that we need 
to study the Pr number is  
the mass density $(\rho)$. In our case, the mass density 
is defined as 
\ba
\rho=2\sum_i m_in_i+m_gn_g, 
\ea
where $m_i~(m_g)$  are the quasi-particle masses 
of the quarks (gluons), generated due to the 
presence of the thermal medium. Apart from it, $n_{eq,i}$ and $n_{eq,g}$ 
are the number densities 
of the quarks and gluons, respectively which 
can be calculated using the 
phase space distribution functions [Eqns.~\eqref{num_quark}
and \eqref{num_gluon}]. The mass density in the absence 
of the magnetic field reads 
\ba
\rho=\frac{1}{\pi^2}\sum_i m_ig_i\int dp~ p^2 f_{\rm eq,i} 
+\frac{1} {2\pi^2}m_gg_g\int dp~ p^2 f_{\mathrm eq,g},
\ea
and for a strongly  magnetized medium, it is
\ba
\rho^B=\frac{1}{2\pi^2}\sum_i m_i g_i|q_iB|\int dp_3 ~f_{\rm eq,i}^B+ 
\frac{1} {2\pi^2}m_gg_g\int dp ~p^2 f_{\mathrm eq,g}.
\ea

\begin{center}
\begin{figure}[H]
\begin{tabular}{cc}
\includegraphics[width=6cm]{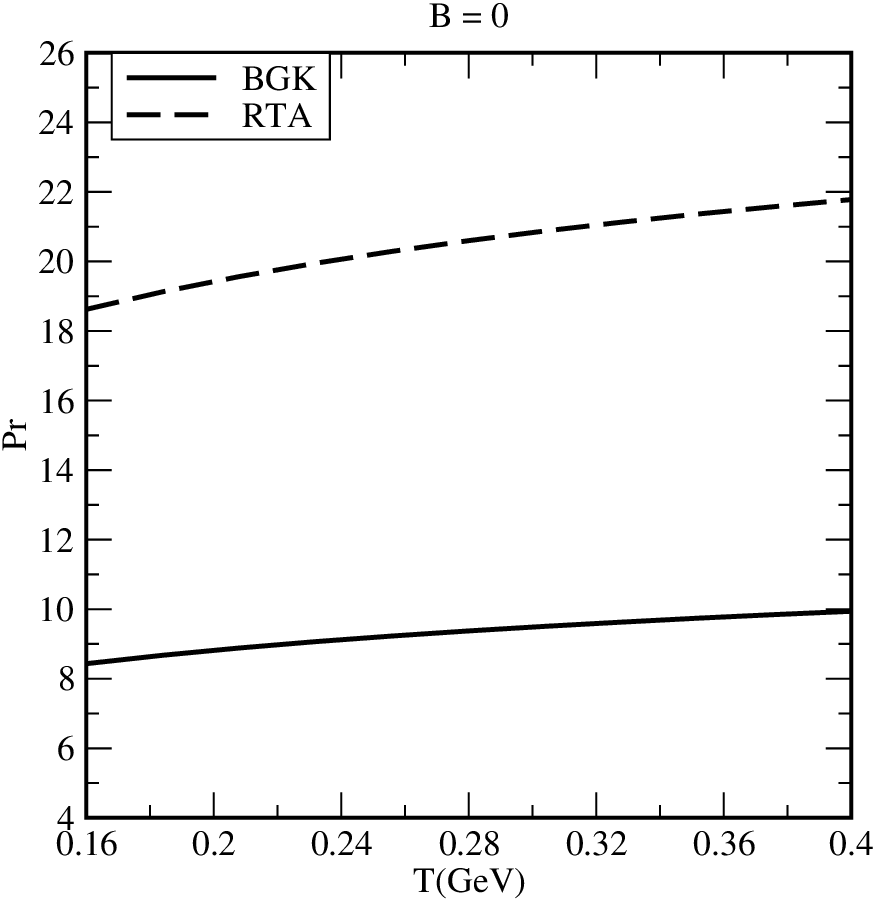}&
\hspace{2.5cm}
\includegraphics[width=6cm]{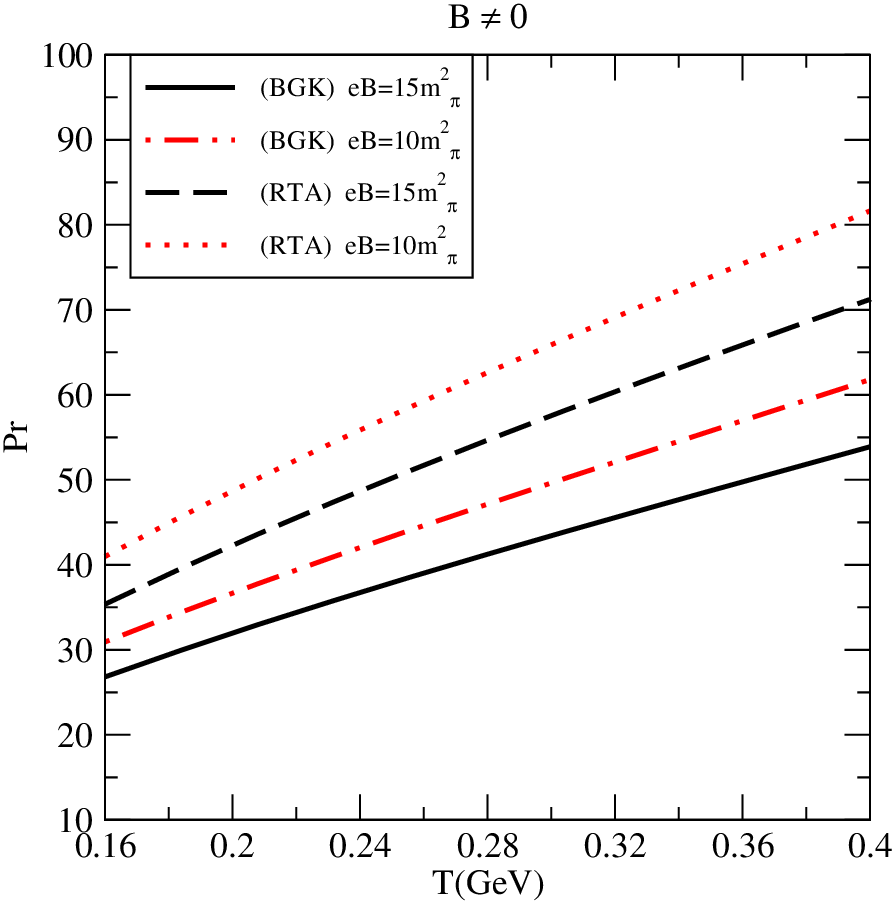}\\
(a)& \hspace{2.5cm}(b)
\end{tabular}
\caption{ Prandtl number as a function of $T$ in the absence (a)
 and in the presence (b) of the strong $B$} 
\label{Pr_num} 
\end{figure}
\end{center}
Now we will focus on our results of the Prandtl number. 
In the left panel of fig.~\ref{Pr_num}, 
we have shown the Pr number as a function of the 
temperature in the absence of $B$. It is found to be increasing 
monotonically with 
$T$ in both BGK and RT collision integrals. 
The magnitude is 
greater than one in both the collision terms 
which means that in a medium
consisting of quarks and gluons, the rate of  momentum
diffusion dominates over that of thermal diffusion. The magnitude 
gets reduced in the BGK collision integral. The reduction 
in the Pr number may lead to the conclusion that instantaneous 
conservation 
 of the particle number enforces less
pronounced momentum transport. We carry out
similar investigations in the presence of the strong $B$ 
 in the right panel of fig.~\ref{Pr_num} and observe
 similar trends with the temperature. There is 
an enhancement  in the magnitude of the 
 Pr number, but its behavior with  the collision integrals 
 is similar {i.e.} RT collision integral dominates over the BGK.
 It decreases with the magnetic field.
Pr number has been evaluated earlier for many systems such as 
dilute atomic Fermi gas~\cite{Braby:PRA82'2010} and 
has been reported to be around $\frac{2}{3}$ at high temperature. In 
case of non-relativistic conformal holographic fluid~\cite{Rangamani:JHEP01'2009}
it has been found to be around $1.0$ and for 
strongly coupled liquid helium, $2.5$. 
\cite{Schafer:RPP72'2009}.    
\subsection{Reynolds number}
 From a hydrodynamic point of view, the Reynolds number has an 
essential significance in determining the 
nature of the flow pattern of a fluid. The small value of 
the Reynolds 
number indicates the laminar flow, while a large one 
tells about the turbulence. It is defined as 
\ba
 {\rm RI}=\frac{Lv}{\eta/\rho},
\ea
 where $L$ refers to the characteristic length while 
$v$ to the relative velocity of the fluid,
respectively. The magnitude of the RI  gives an idea about the 
 kinematic viscosity $\eta/\rho$ of the fluid in comparison to 
 the characteristic
 length and the relative speed. The large value 
 (in case of turbulent flow)
  of the RI corresponds to the 
small magnitude of  $\eta/\rho$  in comparison to the quantity
 $Lv$ of the system.
 \vspace{8mm} 
\begin{center}
\begin{figure}[H]
\begin{tabular}{cc}
\includegraphics[width=6cm]{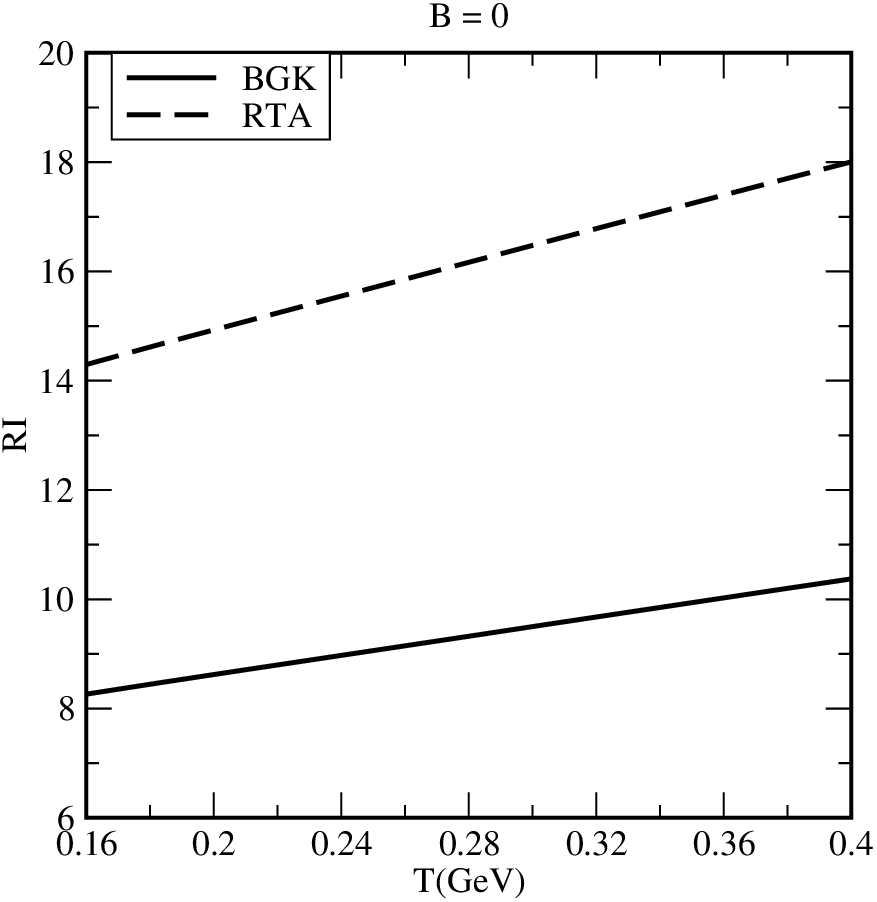}&
\hspace{2.5cm}
\includegraphics[width=6cm]{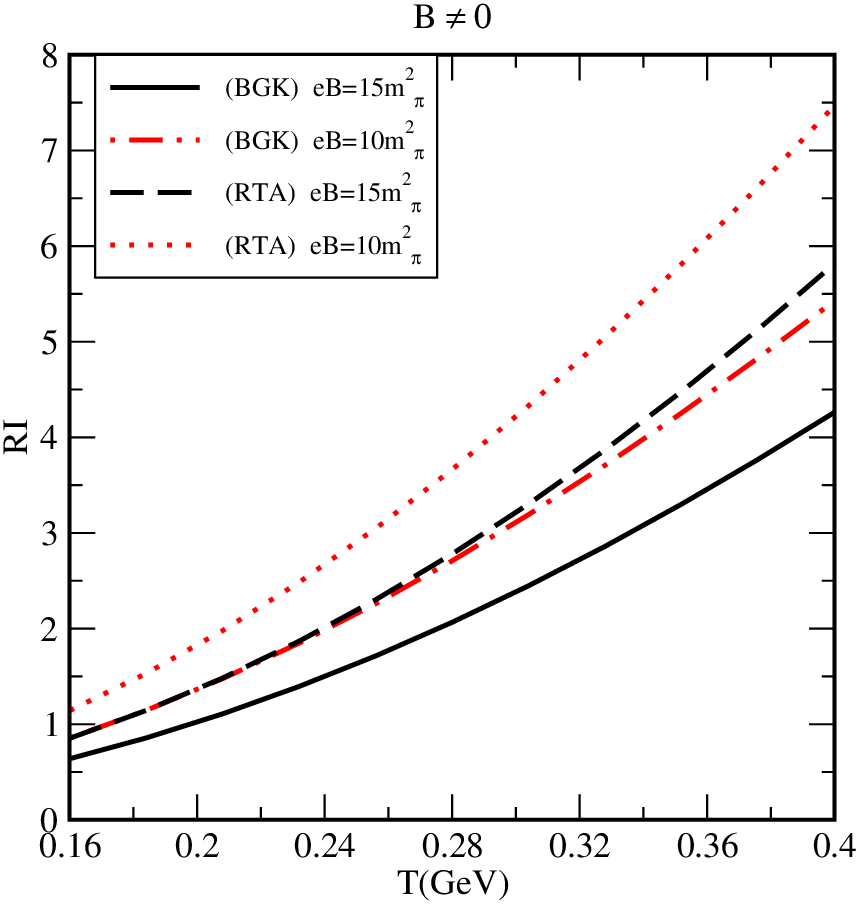}\\
(a)& \hspace{2.5cm}(b)
\end{tabular}
\caption{ Reynolds number as function of $T$ in absence (a)
 and in presence (b) of the strong $B$} 
\label{Reynold} 
\end{figure}
\end{center}
To study the impact of the BGK collision term 
 on  the nature of the flow of the thermal QCD medium, we plot Reynolds 
number in the left panel of Fig.~\ref{Reynold} with  
$T$ in the absence of the magnetic field. RI  increases
 with the temperature and its magnitude
 gets lowered in the BGK type collision integral in 
 comparison to the RT, which indicates 
that instantaneous conservation of 
particle number promotes the laminar nature of the flow. 
The value of the RI is found to be 
around $8-10$ for BGK and $14-18$ for RT in the temperature 
range $160<T<400$ MeV. In a (3 + 1)-dimensional fluid dynamical
model, the value of the RI is estimated in the range
$3-10$ for initial QGP with $\eta/s=$  0.1 \cite{Csernai:PRC85'2012}, 
The holographic model reports its upper bound as 20 
\cite{McInnes:NPB21'2017}. In the
 right panel, we perform  
similar studies in the presence of 
 strong magnetic field. 
The trends are similar to the $B=0$
 case {\em i.e.} it increases with 
$T$ and its magnitude gets reduced
 in BGK term. However, RI shows decreasing trends with 
 increasing $B$. The value of the RI is roughly in the 
 range of $1-8$ as $T$ is varied from $160$ to $400$ MeV
 for the two strengths of the magnetic field {\em i.e.} 
 $eB=15m^2_{\pi}$ and $eB=10m^2_{\pi}$. 
 We have observed that the RI is reduced 
in the presence of strong $B$ as compared to $B=0$ case and this 
reduction is more pronounced in the low $T$ region around
the QCD transition point.
  
 As seen earlier, the temperature
dependence of $\eta$ and $\eta/s$ with 
BGK collision term at $eB=10 m_\pi^2$ shows a similar
behaviour with its counterpart with RT collision term at
$eB=15 m_\pi^2$ in Fig.~\ref{shear} (a) and 
Fig.~\ref{shear_s} (b), respectively. Apart from
the size of the system ($L$) and the 
relative velocity ($v$), Reynolds 
number is obtained from the inverse 
of kinematic viscosity ($\eta/\rho$), 
so RI also shows a similarity between 
BGK predictions at $eB=10 m_\pi^2$ 
and RT predictions at $eB=15 m_\pi^2$.
This similarity  
leads to an ambiguity in the phenomenological
modeling. The BGK collision term should be
prefered over RTA for the extraction of 
phenomenologically relavant quantities from 
experimental data. 

 \subsection{ Momentum diffusion
vs charge diffusion}
The relative competition between the momentum diffusion and charge diffusion 
 can be better understood through a dimensionless ratio
\ba
\gamma=\frac{\eta/s}{\sigma_{el}/T},
\label{gamma_fac} 
\ea
where $\sigma_{el}$ is the electrical conductivity. The 
quarks, unlike gluons, are electrically charged particles, hence,  
 contribute to $\sigma_{el}$,
while both quarks and gluons contribute
 to  the momentum transport, hence,
contribute to the shear viscosity. 
The ratio $\gamma$ gives an idea 
about the relative significance of the matter 
and gluon sector contributions to 
 the momentum and charge diffusion in the hot QCD medium.
 \vspace{8mm}
\begin{center}
\begin{figure}[H]
\begin{tabular}{cc}
\includegraphics[width=6cm]{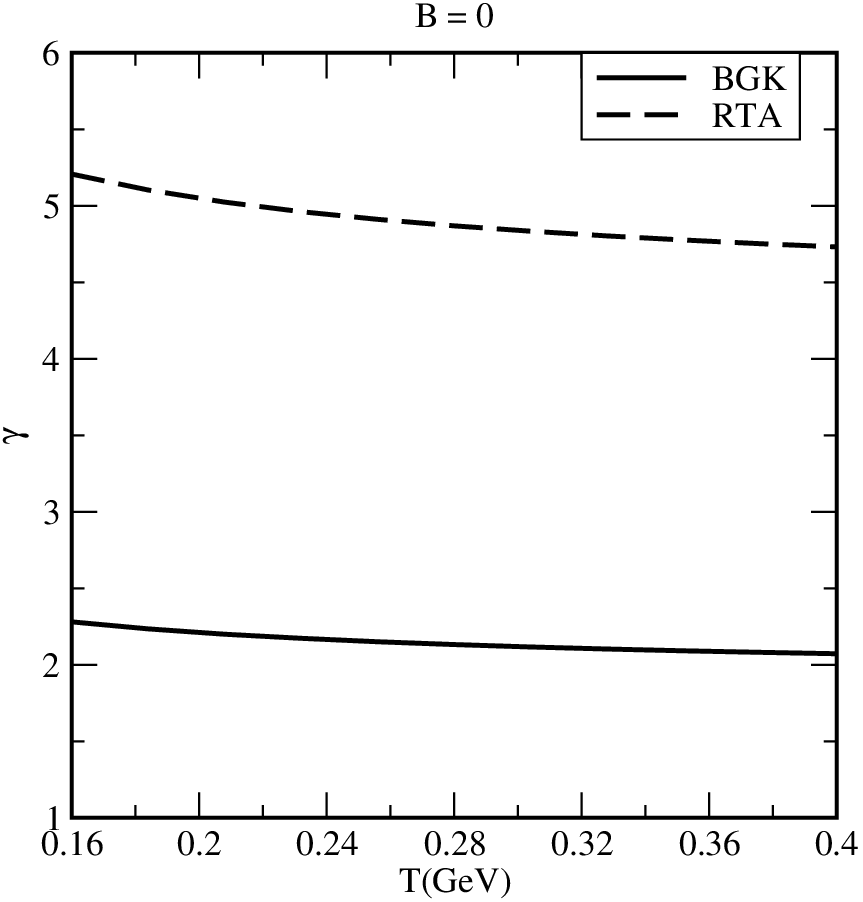}&
\hspace{2.5cm}
\includegraphics[width=6cm]{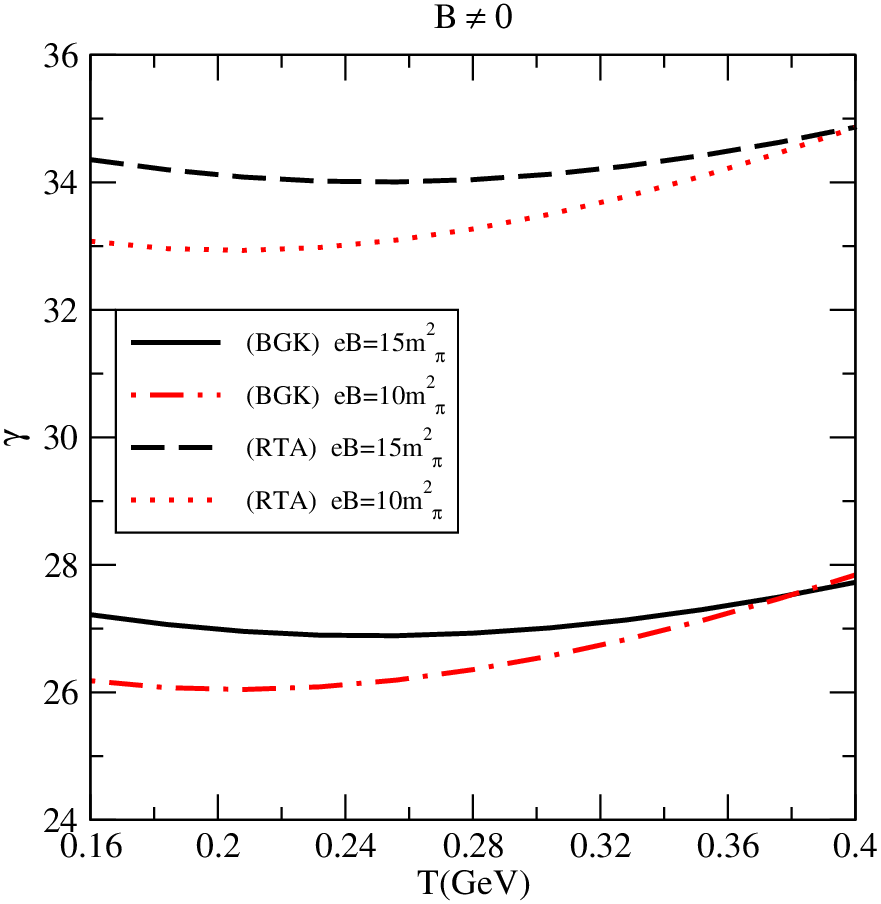}\\
(a)& \hspace{2.5cm}(b)
\end{tabular}
\caption{ Dimensionless ratio $\gamma$ as a function of 
$T$ in absence (a) and in presence (b) of the strong $B$.} 
\label{gamma} 
\end{figure}
\end{center}
In Fig.~\ref{gamma}, we have shown the
 dimensionless ratio $\gamma$ calculated
  in Eqn.~\eqref{gamma_fac}
 as a function of $T$ in the absence of  $B$ (in the 
left panel) and have found that $\gamma$ 
decreases slowly with temperature 
near the crossover point ($\sim 160 )$ 
MeV and then almost remains constant at
higher temperatures. The magnitude of 
the $\gamma$ factor gets 
reduced in the BGK collision integral, which indicates 
 that conservation of particle number enhances the
charge transport in the medium at a greater rate in 
comparison to momentum
transport. In the right panel of Fig.~\ref{gamma}, we have displayed
the ratio $\gamma$ with temperature in the presence of the 
strong $B$ and have observed the reduction of $\gamma$ in BGK 
collision term like the $B=0$ case. The $\gamma$ factor first decreases
with $T$ and then increases. Its magnitude is 
higher when the strength of the 
 magnetic field is $eB=15m^2_{\pi}$ in comparison to  $eB=10m^2_{\pi}$.
\subsection{  Bulk viscosity vs shear viscosity}
The relative competition between the shear viscous  and 
 bulk viscous nature of the fluid can be understood  in terms of the 
ratio $\zeta/\eta$. This ratio has been  calculated for a variety of 
systems in the different frameworks. In the 
case of interacting scalar field~\cite{Horsley:NPB280'1987}, 
$\zeta/\eta$ has been found to be
around $15\left(\frac{1}{3}-c_s^2\right)^2$ (where $c_s$ refers
to the speed of sound in the medium). For
hot QCD medium using the perturbation theory, it is nearly equal to
 $15\left(\frac{1}{3}-c_s^2\right)^2$~
\cite{Arnold:PRD74'2006}, and 
for strongly coupled gauge plasma, $2\left(\frac{1}{3}-c_s^2\right)^2$~\cite{Buchel:PLB663'2008}. In  holographic model~\cite{Buchel:NPB820'2009},
the ratio $\zeta/\eta$ is found to be less than $0.5$
at high temperatures and it is around $0.6$ in the 
phase transition region. In another 
attempt~\cite{Gursoy:JHEP0912'2009}, 
the authors notice that $\zeta$ is smaller than 
$\eta$ at  high temperatures but acquires higher values around
 critical temperature ($T_c$) of the phase transition.
In the case of quasi-gluon plasma~\cite{Bluhm:PLB709'2012},
 the ratio $\zeta/\eta$ behaves
like that found using the  perturbative QCD 
at  higher  temperatures (above $1.5~T_c$) but 
for  $T$ around $1.02~T_c$,  
nonperturbative effects become significant.
\vspace{8mm}
\begin{center}
\begin{figure}[H]
\begin{tabular}{cc}
\includegraphics[width=6cm]{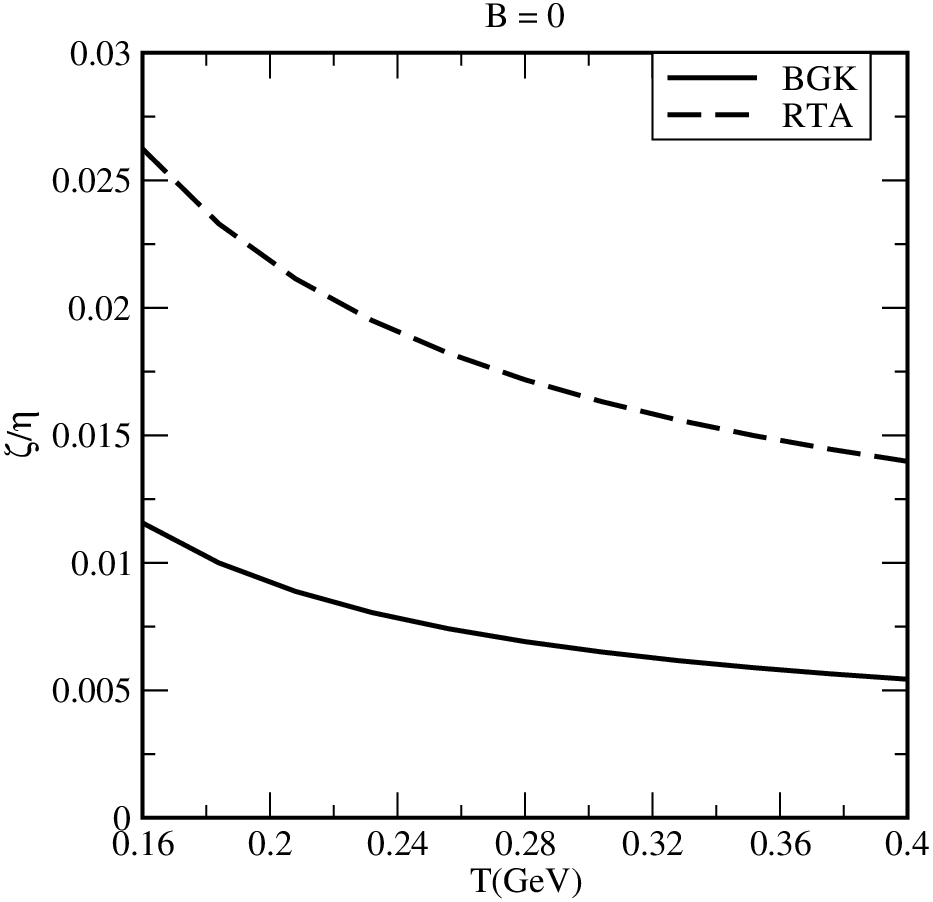}&
\hspace{2.5cm}
\includegraphics[width=6cm]{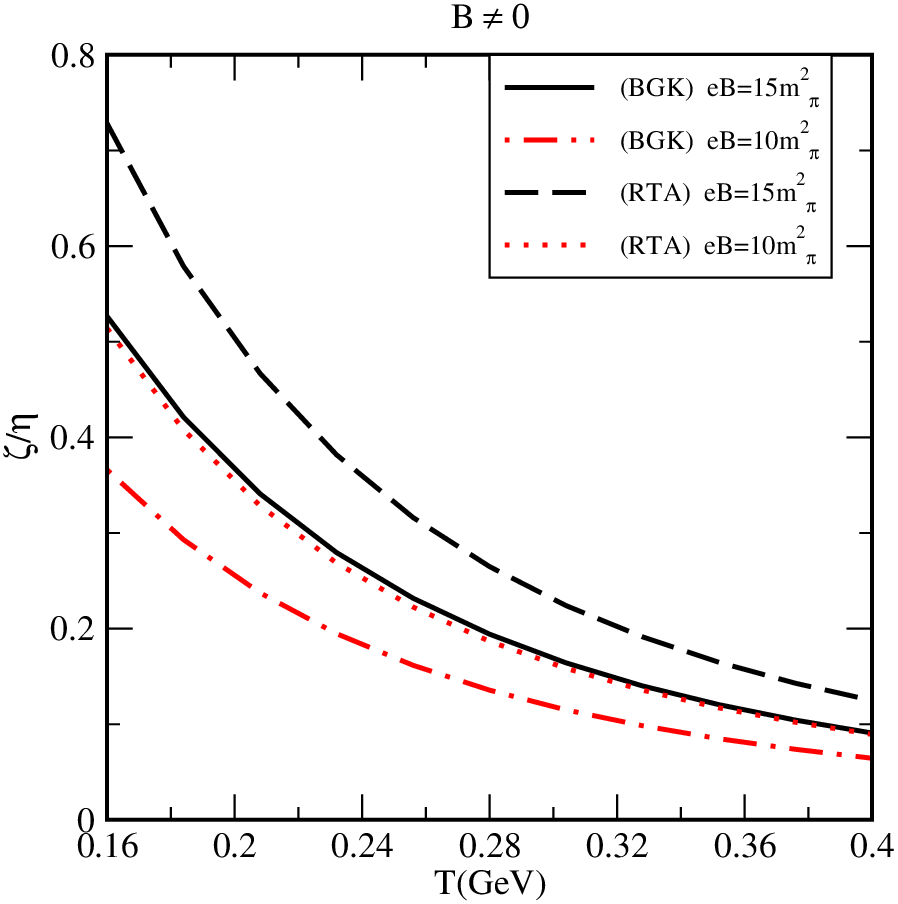}\\
(a)& \hspace{2.5cm} (b)
\end{tabular}
\caption{  Bulk to shear viscosity $(\zeta/\eta)$ ratio as a function 
of $T$ in absence (a)
 and in presence (b) of the strong $B$} 
\label{eta_zeta} 
\end{figure}
\end{center}
Now we will explore the 
ratio $\zeta/\eta$ for our case.
 In fig.~\ref{eta_zeta}, 
the variation of the ratio 
$\zeta/\eta$ with temperature has been 
shown. It decreases with $T$ and gets 
reduced in BGK collision term in the 
absence [Fig.~\ref{eta_zeta} (a)] as well as  in
the presence  of strong $B$ [Fig.~\ref{eta_zeta} (b)]. 
In the absence of $B$, the magnitude of the 
ratio $\zeta/\eta$ remains less than one 
in the temperature range 
 $160-400$ MeV that corresponds to the 
dominance of $\eta$ over $\zeta$. In
SMF, the ratio $\zeta/\eta$ gets enhanced slightly,
 but the magnitude is 
still, less than one and the RT 
collision term dominates over BGK.
\section{Conclusion}\label{5}
 To conclude, we have examined the relative 
behavior of the transport coefficients 
of the thermal QCD medium in BGK and RT collision 
integrals. This study helps us in probing 
the various salient features of the collision integrals 
and their impact on the transport phenomenon.
For that purpose, we have evaluated shear and bulk viscosities
using the relativistic Boltzmann transport equation
where collisional aspects of the medium have been incorporated
with the help of BGK type collision integral, which exhibits  
an improvement
over RT enforcing the conservation of particles in the medium. 
Then using $\eta$ and $\zeta$, we
have studied the ratios $\eta/s$ and $\zeta/s$ 
to get an idea about the 
ideal  and conformal nature of the fluid. 
We further explored the relative significance of the various transport coefficients through  Prandtl number, Reynolds number, 
factor $\gamma$
and ratio $\zeta/\eta$.
 Both $\eta$ and $\zeta$ have been found to be increasing 
 with the temperature in both the 
 collision integrals but
the magnitude of $\eta$ ($\zeta$) 
gets enhanced (reduced) in BGK collision
term in comparison to RT. As a result, 
$\eta/s$ gets enhanced whereas $\zeta/s$ reduces. The 
magnitude of other derived coefficients Pr, RI, $\gamma$ and 
ratio $\zeta/\eta$ also 
gets reduced. Apart from it,
the ratio $\eta/s$ ($\zeta/s$) is minimum (maximum) near $T_c$. 
Further, we studied the impact of strong $B$ on the collision integral 
and subsequently on 
the momentum transport in the medium where quark dynamics
is restricted in only 1-D {\em i.e.} along the direction 
of the $B$. The abovementioned transport coefficients
show similar trends  with respect to the collision integral as
 $B=0$ case  except for ratio $\zeta/s$, which is almost
identical in both the collision integrals. The
 strong $B$ flips the $T$ dependence of $\zeta$ and $\eta/s$
 which now shows decreasing trends. The ratio $\eta/s$ ($\zeta/s$) 
becomes smaller (larger). We also
see enhancement of Pr, $\gamma$, and $\zeta/\eta$. 
Enhancement of Pr
number leads to the conclusion that in a strong $B$, 
the sound attenuation is mostly controlled by $\eta$. 
In strong $B$, RI gets reduced, which indicates that
strong B adds to the laminar nature of the flow. 
We see that different collision integral 
gives different values of the transport coefficients 
which are  experimentally measurable quantities. Thus 
using this one-to-one correspondence, we can sense
the collision integral   
responsible for the equilibration of the medium.   
In this study, the hot QCD medium effects have been 
incorporated via dispersion relations  wherein the masses of the 
quasi-partons are parameterized according to a  
thermodynamically consistent quasi-particle model. 
We calculate the medium-generated thermal masses of the partons 
 by taking the poles 
of the resummed  propagators calculated under the framework of  perturbative
thermal QCD  with a strong $B$ in the background.          
\section*{Acknowledgements}
S. A. Khan is thankful to Shubhalaxmi Rath for useful
discussions  during the course of the work. 
\appendix
\appendixpage
\addappheadtotoc
\begin{appendix}
\renewcommand{\theequation}{A.\arabic{equation}}
\section{Derivation of the left hand side in equation \eqref{BGk_left}}
 The  BGK collision term (right hand side) of~ \eqref{RBTE_quark} 
is given as
\ba
C[f]&=&-p^{\mu}u_{\mu}\nu_i \left(f_i-n_{{}_i}
n_{\mathrm eq,i}^{-1} f_{\mathrm eq,i}
\right),\nonumber\\
&=&-p^{\mu}u_{\mu}\nu_i\left(f_i-
\frac{g_i \int_p (f_{\mathrm eq,i}+\delta f_i)}{n_{\mathrm eq,i}}
f_{\mathrm eq,i}\right),\nonumber\\
&=&-p^{\mu}u_{\mu}\nu_i\left(f_i-\frac{(g_i \int_p f_{\mathrm eq,i}
+g_i\int_p\delta f_i)}{n_{\mathrm eq,i}}f_{\mathrm eq,i}\right),\nonumber\\
&=&-p^{\mu}u_{\mu}\nu_i\left(\delta f_i-g_i n_{\mathrm eq,i}^{-1}
f_{\mathrm eq,i}\int_p\delta f_i\right).
\ea 
\renewcommand{\theequation}{B.\arabic{equation}}
\section{ Quark self energy in imaginary time formalism}
We can approximate the exponential factor  in \eqref{quark_prop}
as $e^{-k_{\perp}/|q_iB|}\approx 1$, since 
the transverse component of the quark momentum is 
almost negligible {\em i.e.} $k_{\perp}\approx 0$.
The quark self energy \eqref{quark_self} 
 takes the form
\begin{eqnarray}
\Sigma(p_\parallel) &=& \frac{2g^2}{3\pi^2}
|q_iB|T\sum_n\int dk_z\frac{\left[\left(1+\gamma^0\gamma^3\gamma^5\right)\left(\gamma^0k_0
-\gamma^3k_z\right)-2m_i\right]}{\left[k_0^2-
\omega^2_k\right]\left[(p_0-k_0)^2-\omega_{pk}^2\right]}\nonumber \\ 
&=& \frac{2g^2|q_iB|}{3\pi^2}\int dk_z
\left[(\gamma^0+\gamma^3\gamma^5)L^1-(\gamma^3+\gamma^0\gamma^5)k_zL^2\right]
\label{quark_self1}
,\end{eqnarray}
where $\omega_k^2=k_z^2+m_i^2$, $\omega_{pk}^2=(p_z-k_z)^2$
and $L^1$ and $L^2$ are the frequency sums needed to evaluate 
the self energy which can be written as 
\begin{eqnarray}
L^1=T\sum_nk_0~\frac{1}{\left[k_0^2-\omega_k^2\right]}
\frac{1}{\left[(p_0-k_0)^2-\omega_{pk}^2\right]}, 
\end{eqnarray}
and
\begin{eqnarray}
L^2=T\sum_n\frac{1}{\left[k_0^2-\omega_k^2\right]}
\frac{1}{\left[(p_0-k_0)^2-\omega_{pk}^2\right]},
\end{eqnarray}
respectively. After performing the frequency sums,  
the self energy \eqref{quark_self1} takes the form
\begin{eqnarray}
\Sigma(p_\parallel)&=&\frac{g^2|q_iB|}{3\pi^2}\int \frac{dk_z}{\omega_k}
\left[\frac{1}{e^{\beta\omega_k}-1}+\frac{1}{e^{\beta\omega_k}+1}\right]\nonumber\\
&&\times\left[\frac{\gamma^0p_0}{p_\parallel^2}+\frac{\gamma^3p_z}{p_\parallel^2}+\frac{\gamma^0\gamma^5p_z}{p_\parallel^2}+\frac{\gamma^3\gamma^5p_0}{p_\parallel^2}\right],
\end{eqnarray}
which can further be simplified after integration over  $k_z$ as 
\begin{eqnarray}
\Sigma(p_\parallel)=\frac{g^2|q_iB|}{3\pi^2}\left[\frac{\pi T}{2m_i}-\ln(2)\right]\left[\frac{\gamma^0p_0}{p_\parallel^2}+\frac{\gamma^3p_z}{p_\parallel^2}+\frac{\gamma^0\gamma^5p_z}{p_\parallel^2}+\frac{\gamma^3\gamma^5p_0}{p_\parallel^2}\right]
\label{quark_self2}
.\end{eqnarray}
\end{appendix}

\end{document}